\documentclass[16pt]{article}
\usepackage{amscd}
\usepackage{bbm}
\usepackage{bm}
\usepackage{mathrsfs}
\usepackage{amssymb}
\usepackage{graphics}
\usepackage{graphicx}
\usepackage{amsfonts}
\usepackage{amsmath}
\usepackage{array}
\usepackage{booktabs}

\usepackage[numbers,sort&compress]{natbib}
\begin{document}
\date{}
\title{Dynamics of the semi-discrete Gardner equation under two types of non-vanishing boundary conditions: heteropolar solitons and kinks}
\author{Jia-Xue Niu,  Yan-Nan Zhao, Rui Guo$\thanks{Corresponding author:
gr81@sina.com}$, Jian-Wen Zhang
\\
\\{\em
School of Mathematics, Taiyuan University  of} \\
{\em Technology, Taiyuan 030024, China}
} \maketitle

\begin{abstract}
In this work, we will use inverse scattering transform to study the semi-discrete Gardner equation under two types of non-vanishing boundary conditions, and investigate two interesting nonlinear waves in the presence of discrete spectrum, namely heteropolar solitons and kinks. When $u_n\rightarrow -\frac{a}{2b}$ as $n\rightarrow \pm \infty $, this is a symmetric boundary condition, for which the heteropolar solitons, i.e., two kinds of single soliton solutions with different polarities will be obtained. If considering two sets of discrete eigenvalues, there will be two types of soliton collisions, head-on and overtaking collision, depending on the position of discrete spectrum. Interestingly, the energy gathered at the moment of collision with different polarities, producing the so-called rogue wave phenomenon with a large amplitude more than twice the background, and its generation mechanism is briefly analyzed. When $u_n\rightarrow \frac{c_{\pm}\sqrt{ a^2+4b }-a}{2b}$ as $n\rightarrow \pm \infty $, the kink, i.e., the undercompressive dispersive shock wave, will be obtained under the specific step-like boundary condition.

\vspace{5mm}\noindent\emph{Keywords}: Semi-discrete Gardner equation; Heteropolar soliton; Kink; Inverse scattering transform; Multi-poles solution; Non-vanishing boundary condition;
\end{abstract}
\vspace{7mm}\noindent\textbf{1  Introduction}
\hspace*{\parindent}
\renewcommand{\theequation}{1.\arabic{equation}}

Due to the uneven distribution of physical quantities such as temperature, salinity and concentration, the fluid density is stratified with spatial position (usually vertical direction). This fluid system, i.e., stratified fluid, is common in natural environments such as the ocean and atmosphere~\cite{ck1}. Internal wave is a kind of nonlinear wave that appears at the interface between different layers, including solitary wave, rogue wave, etc. Generally, the Korteweg-de Vries (KdV) equation is the basic model for the description of nonlinear internal waves~\cite{ck2}. However, in a two-layer fluid, the description of nonlinear internal waves requires high-order nonlinear effects, then the modified Korteweg-de Vries (mKdV) equation~\cite{ck3} was introduced. As the system that can describe the dynamics of internal waves for an arbitrary vertical stratification of the density field~\cite{ck4,ck5,ck6}, the Gardner equation~\cite{ck7}
\begin{equation}\tag{1.1}
u_t+6 uu_x-6\alpha u^2u_x+u_{xxx}=0
\end{equation}
which contains both quadratic and cubic nonlinear terms, has been shown to describe nonlinear wave effects in a number of physical contexts including plasma physics~\cite{ck8,ck9}, stratified fluid flows~\cite{ck10}, and quantum fluid dynamics~\cite{ck11}. In the case where the amplitudes of the left and right far-field are equal, Eq.~(1.1) admits soliton solutions~\cite{ck12} with different polarities due to its invariance for the transformation: $u\rightarrow \frac{1}{\alpha}-u$. And the dynamical properties of soliton collisions for Eq.~(1.1) are analyzed in Ref.~\cite{ck13}, it has been shown that the soliton polarity affects radically the result of the interaction between the solitons, the bright and dark solutions of Eq.~(1.1) can model internal rogue waves in three-layer fluids~\cite{ck14,ck15}. In addition, because Eq.~(1.1) contains cubic nonlinear terms for which the possible signs correspond to different fluid density stratification profiles~\cite{ck16}, it can exhibit non-convex properties, which leads to the existence of kink solutions~\cite{ck12,ck17} relative to its reduction, the convex KdV equation. Generally speaking, the kink can be obtained through the limit case of periodic waves\cite{ck12,ck18,ck19}, it is regarded as the uncompressive dispersive shock wave and can describe the sudden change of medium in many physical environments, including dam problem~\cite{ck20}, piston problem~\cite{ck21} and the propagation of stress wave during earthquake.

It should be noted that the collisions of heteropolar solitons and the existence of kinks are accompanied by different boundary conditions, which were not emphasized in the above analysis of Eq.~(1.1). Naturally, it is worth exploring whether similar properties can be derived from specific boundary conditions for its discrete counterpart, the semi-discrete Gardner equation~\cite{ck22}
\begin{align}
\frac{d}{dt}u_n&+\left( 1-au_n-bu_{n}^{2} \right) \left[ a\left( u_{n+1}^{2}+u_{n+1}u_{n+2}+u_nu_{n+1}-u_nu_{n-1}-u_{n-1}u_{n-2}-u_{n-1}^{2} \right) \right.\notag\\
&\left.+b\left( u_{n+1}^{2}u_n+u_{n+1}^{2}u_{n+2}-u_{n-1}^{2}u_{n-2}-u_{n-1}^{2}u_n \right) +u_{n-2}+2u_{n+1}-2u_{n-1}-u_{n+2} \right] =0.\tag{1.2}
\end{align}
As is known to all, the discrete nonlinear integrable system is an important branch of integrable system theory~\cite{ck23,ck24,ck25,ck26,ck30,ck42,ck43}. It is obtained by discretizing time or space and transforming the original problem into a system of algebraic equations, which can be directly solved through numerical algorithms. Moreover, in scenarios where certain continuity assumptions fail, such as describing atomic motion, simulating particulate matter, and certain discrete systems, discrete nonlinear integrable systems become particularly necessary. As an important discrete system, Eq.~(1.2) is integrable through the Lax pair
\begin{equation}\tag{1.3}
\psi _{n+1}=U_n\psi _n,\ \ \ \frac{d\psi _n}{dt}=V_n\psi _n,
\end{equation}
where
$$
U_n=\left( \begin{matrix}
	\lambda&		\frac{2bu_n+a}{\sqrt{\sigma \left( a^2+4b \right)}}\\
	\sigma \frac{2bu_n+a}{\sqrt{\sigma \left( a^2+4b \right)}}&		\lambda ^{-1}\\
\end{matrix} \right)
$$
$$
V_n=\left( \begin{matrix}
	K_{n}^{\left( 1 \right)}\left( \lambda \right)&		L_{n}^{\left( 1 \right)}\left( \lambda \right)\\
	\sigma L_{n}^{\left( 1 \right)}\left( \lambda ^{-1} \right)&		K_{n}^{\left( 1 \right)}\left( \lambda ^{-1} \right)\\
\end{matrix} \right) +\left( \frac{a^2+4b}{4b} \right) ^2\left( \begin{matrix}
	K_{n}^{\left( 2 \right)}\left( \lambda \right)&		L_{n}^{\left( 2 \right)}\left( \lambda \right)\\
	\sigma L_{n}^{\left( 2 \right)}\left( \lambda ^{-1} \right)&		K_{n}^{\left( 2 \right)}\left( \lambda ^{-1} \right)\\
\end{matrix} \right)
$$
with
$$
K_{n}^{\left( 1 \right)}\left( \lambda \right) =\frac{3a^2\left( a^2+4b \right)}{8b^2}\lambda ^2-\frac{3a^2\left( 2bu_n+a \right) \left( 2bu_{n-1}+a \right)}{8b^2}-\frac{3a^2}{2b},
$$
$$
L_{n}^{\left( 1 \right)}\left( \lambda \right) =\frac{3a^2\left( a^2+4b \right)}{8b^2}\left( \frac{2bu_n+a}{\sqrt{\sigma \left( a^2+4b \right)}}\lambda +\frac{2bu_{n-1}+a}{\sqrt{\sigma \left( a^2+4b \right)}}\lambda ^{-1} \right) ,
$$
\begin{align}
K_{n}^{( 2 )}( \lambda ) &=\frac{1}{a^2+4b}[2( 2bu_{n-1}+a ) ( 2bu_n+a )-( 2bu_{n-2}+a ) ( 2bu_n+a )-( 2bu_{n+1}+a ) ( 2bu_{n-1}+a )]\notag\\
&+\frac{1}{(a^2+4b)^2}\left[\left( 2bu_n+a \right)^2 \left( 2bu_{n-1}+a \right)^2+\left( 2bu_n+a \right) \left( 2bu_{n-1}+a \right) ^2\left( 2bu_{n-2}+a \right)\right.\notag\\
&\left.+\left( 2bu_n+a \right) ^2\left( 2bu_{n-1}+a \right) \left( 2bu_{n+1}+a \right)\right]-\frac{3}{2}-(\lambda ^2-\lambda ^{-2})\frac{\left( 2bu_n+a \right) \left( 2bu_{n-1}+a \right)}{a^2+4b}+2\lambda ^{-2}+\frac{\lambda ^4}{4}-\frac{3}{4\lambda ^4}, \notag\\
L_{n}^{\left( 2 \right)}\left( \lambda \right) &=\frac{1}{\sqrt{\sigma \left( a^2+4b \right)}}\left[ \lambda ^3\left( 2bu_n+a \right) +\lambda \left( 2bu_{n+1}-4bu_n-a-2 \left( bu_{n-1}+bu_{n+1}+a \right)\frac{(2bu_n+a)^2}{ a^2+4b} \right) \right.\notag\\
&\left.+\lambda ^{-1}\left( 2bu_{n-2}-4bu_{n-1}-a-2\left( bu_{n-2}+bu_n+a \right)  \frac{(2bu_{n-1}+a)^2}{ a^2+4b }  \right) +\lambda ^{-3}\left( 2bu_{n-1}+a \right)  \right].\notag
\end{align}
The Darboux transformations for Eq.~(1.2) has been constructed, and the exact solutions including table-top solitons, ordinary soliton, rational soliton and their hybrid solutions within the non-zero seed background have been obtained~\cite{ck27,ck28}. In order to construct solutions from boundary conditions and analyze the relationship between potential and spectrum, in this paper, we will study Eq.~(1.2) by the inverse scattering transform(IST).

In the framework of IST, the spectral data determined by the boundary conditions act as a bridge. The simplest one is the vanishing boundary condition, under which the $N$-soliton solution of Eq.~(1.1) has been studied~\cite{ck29}. Compared with the continuous equation, the spectral problems of discrete equations show more complex symmetry properties. In recent years, IST has been successfully used to solve the Ablowitz-Ladik(AL) equation under vanishing boundary conditions~\cite{ck30,ck31}, and the degenerate results of Eq.~(1.2), namely the discrete KdV~\cite{ck32} and the discrete mKdV equation~\cite{ck33}, have also been studied. Noting that the exact solution obtained by IST is usually related to the poles of the reflection coefficients. Scholars have analyzed the AL equation~\cite{ck31}, the discrete mKdV equation~\cite{ck33} and the discrete sine-Gordon equation~\cite{ck34} with high-order poles. Further more, crossing the obstacle of multi-valued Riemannian surface, the non-vanishing boundary condition have been considered for the AL equation~\cite{ck35,ck36,ck37,ck38}. The successful application of IST to discrete equations under symmetric boundary conditions allows us to use this method to study the collision of heteropolar solitons for Eq.~(1.2) in Section 2, a phenomenon that has been discovered in its continuous counterpart Eq.~(1.1). On the other hand, the study of periodic solutions for continuous equations has been ongoing for many years, but due to the fact that the explicit form of periodic solutions for discrete equations is often unavailable, research on them is still limited~\cite{ck39,ck40}. This leads to few reported kink solutions for discrete equations. Thus, in this work, we will study a special step-like boundary condition from another perspective within the framework of IST, which can represent the inconsistency of left and right far-field amplitudes, to explore whether there exists the kink solution for Eq.~(1.2), which has been achieved for the continuous defocusing mKdV equation~\cite{ck41}.

The structure of the present paper is as follows. For simplicity, in Section 2, we will consider a relatively simple non-vanishing boundary condition, strictly analyze its spectral problem, convert the information in the boundary condition into scattering data, study the inverse problem corresponding to the one- and two-order poles respectively, derive the exact expression of soliton solutions with different polarities, and analyze the collision problem between solitons. Further, a specific step-like boundary condition will be considered in Section 3 to study the kink solutions, overcoming the multivaluability caused by the boundary condition. Finally, in Section 4, we will summarize the full text and put forward the problems that can be further studied in the future.

\vspace{5mm} \noindent\textbf{2  The symmetric boundary condition, \bm{$u_n$} tends to \bm{$-\frac{a}{2b}$} as \bm{$n\rightarrow \pm \infty $}}
\hspace*{\parindent}
\renewcommand{\theequation}{2.\arabic{equation}}\setcounter{equation}{0}

We first consider the case of $u_n\rightarrow -\frac{a}{2b}$ when $n\rightarrow \pm \infty $, for which
\begin{equation}\tag{2.1}
U_n\rightarrow U_{\pm}=\left( \begin{matrix}
	\lambda&		0\\
	0&		\lambda ^{-1}\\
\end{matrix} \right) ,\ \
 V_n\rightarrow V_{\pm}=\left( \begin{matrix}
	\omega _1&		0\\
	0&		\omega _2\\
\end{matrix} \right)
\end{equation}
for
$$\omega_1=\frac{1}{16b^2}(-24a^2b+6a^2\left( a^2+4b \right) \lambda ^2+\frac{\left( a^2+4b \right) ^2\left( \lambda ^2-1 \right) ^3\left( \lambda ^2+3 \right)}{4\lambda ^4}),
$$
$$\omega_2=\frac{1}{16b^2}(-24a^2b+\frac{6a^2\left( a^2+4b \right)}{\lambda ^2}-\frac{\left( a^2+4b \right) ^2\left( \lambda ^2-1 \right) ^3\left( 3\lambda ^2+1 \right)}{4\lambda ^4}).
$$

\vspace{5mm}\noindent\textbf{2.1 Jost eigenfunctions and scattering data }

Ignoring the impact of time, we introduce the Jost eigenfunctions satisfying the following boundary conditions
$$
\varPhi _n\left( \lambda \right) \rightarrow\left( \begin{matrix}
	\lambda ^n&		0\\
	0&		\lambda ^{-n}\\
\end{matrix} \right) \ \ \ n\rightarrow -\infty ,\ \ \varPsi _n\left( \lambda \right) \rightarrow\left( \begin{matrix}
	\lambda ^n&		0\\
	0&		\lambda ^{-n}\\
\end{matrix} \right) \,\,\,\,\,\,n\rightarrow +\infty ,
$$
which need to be modified as
\begin{equation}\tag{2.2}
M_n=\left( \begin{matrix}
	\lambda ^{-n}&		0\\
	0&		\lambda ^n\\
\end{matrix} \right) \varPhi _n\left( \lambda \right) \rightarrow I,\ N_n=\left( \begin{matrix}
	\lambda ^{-n}&		0\\
	0&		\lambda ^n\\
\end{matrix} \right) \varPsi _n\left( \lambda \right) \rightarrow I,
\end{equation}
for the convenience of subsequent analysis.\\ \\
\textbf{Proposition 2.1} {\it It can be affirmed that the modified eigenfunctions defined in Eqs.~(2.2) that satisfy specific boundary conditions have the following properties:

(1) Due to the uniqueness of the solution for scattering problem, matrix $\varPhi _n\left( \lambda \right)$ and $\varPsi _n\left( \lambda \right)$ are linearly correlated i.e., there exists scattering matrix $S(\lambda)$ such that $\varPhi _n\left( \lambda \right)=\varPsi _n\left( \lambda \right)S(\lambda)$ with $
S\left( \lambda \right) =\left( \begin{matrix}
	a\left( \lambda \right)&		\overline{b}\left( \lambda \right)\\
	b\left( \lambda \right)&		\overline{a}\left( \lambda \right)\\
\end{matrix} \right)
$. The zeros of $a(\lambda)$ and $\bar{a}(\lambda)$ define the discrete spectrum, corresponding to the solitons.

(2) Analyticity. $M_{n,1}\left( \lambda \right)$ and $N_{n,2}\left( \lambda \right)$ are holomorphic in $D_{out}$ and continuous to $\left| \lambda \right|=1$(The subscript $i$ represents the $i$-th column of the matrix, $i$=1,2), while $M_{n,2}\left( \lambda \right)$ and $N_{n,1}\left( \lambda \right)$ are holomorphic in $D_{in}$ and continuous to $\left| \lambda \right|=1$ for the two complementary regions $
D_{out}=\left\{ \lambda \in \mathbb{C}\ |\ \left| \lambda \right|>1 \right\} $ and $D_{in}=\left\{ \lambda \in \mathbb{C}\,\,|\,\,\left| \lambda \right|<1 \right\}
$. Thus the scattering coefficients $a(\lambda)$ is analytic in $D_{out}$ with $J$ zeros $\lambda_{j}$, $j=1,...,J$; $\overline{a}(\lambda)$ is analytic in $D_{in}$ with $J$ zeros $\overline{\lambda}_{j}$, $j=1,...,J$; $b(\lambda)$ and $\overline{b}(\lambda)$ are only analytic on $\left| \lambda \right|=1$.

(3) Asymptotic behaviors.\\
For $\lambda \rightarrow \pm \infty$,
\begin{equation}\tag{2.3}
M_{n,1}=\left( \begin{array}{c}
	1+O\left( \lambda ^{-2},even \right)\\
	\lambda ^{-1}\sigma \frac{2bu_{n-1}+a}{\sqrt{\sigma (a^2+4b)}}+O\left( \lambda ^{-3},odd \right)\\
\end{array} \right) ,\ \ N_{n,2}=\left( \begin{array}{c}
	-\lambda ^{-1}c_{n}^{-1}\frac{2bu_n+a}{\sqrt{\sigma (a^2+4b)}}+O\left( \lambda ^{-3},odd \right)\\
	c_{n}^{-1}+O\left( \lambda ^{-2},even \right)\\
\end{array} \right) ,
 \end{equation}
\begin{equation}\tag{2.4}
 a\left( \lambda \right) =1+O\left( \lambda ^{-2},even \right)  .
\end{equation}
For $\lambda \rightarrow 0$,
\begin{equation}\tag{2.5}
M_{n,2}=\left( \begin{array}{c}
	\lambda \frac{2bu_{n-1}+a}{\sqrt{\sigma (a^2+4b)}}+O\left( \lambda ^3,odd \right)\\
	1+O\left( \lambda ^2,even \right)\\
\end{array} \right) ,\,\,\,\,N_{n,1}=\left( \begin{array}{c}
	c_{n}^{-1}+O\left( \lambda ^2,even \right)\\
	-\lambda c_{n}^{-1}\sigma \frac{2bu_n+a}{\sqrt{\sigma (a^2+4b)}}+O\left( \lambda ^3,odd \right)\\
\end{array} \right) ,
\end{equation}
\begin{equation}\tag{2.6}
\bar{a}\left( \lambda \right) =1+O\left( \lambda ^2,even \right) .
\end{equation}

(4) On the one hand, we can build the symmetry for modified eigenfunctions from $D_{out}$ to $D_{in}$. For $\lambda \in D_{in}$,
\begin{equation}\tag{2.7}
M_{n,2}\left( \lambda \right) =\left( \begin{matrix}
	0&		\pm 1\\
	1&		0\\
\end{matrix} \right) M_{n,1}\left( \lambda ^{-1} \right) ,\ \ N_{n,1}\left( \lambda \right) =\left( \begin{matrix}
	0&		1\\
	\pm 1&		0\\
\end{matrix} \right) N_{n,2}\left( \lambda ^{-1} \right) ,
\end{equation}
\begin{equation}\tag{2.8}
\overline{a}\left( \lambda \right) =a\left( \lambda ^{-1} \right) ,\ \ \overline{b}\left( \lambda \right) =\pm b\left( \lambda ^{-1} \right) ,
\end{equation}
in which "$\pm $" corresponding to $\sigma =\pm 1$. On the other hand, based on the parity of the modified eigenfunctions expansion, we obtained the symmetry under the mapping $\lambda \rightarrow -\lambda$ as follows
\begin{equation}\tag{2.9}
M_n\left( -\lambda \right) =\left( \begin{matrix}
	1&		-1\\
	-1&		1\\
\end{matrix} \right) M_n\left( \lambda \right) ,\ \ N_n\left( -\lambda \right) =\left( \begin{matrix}
	1&		-1\\
	-1&		1\\
\end{matrix} \right) N_n\left( \lambda \right).
\end{equation}
}

According to the definition of scattering matrix, in which the elements can be represented by the Wronskian of modified eigenfunctions
\begin{equation}\tag{2.10a}
a\left( \lambda \right) =x_nWr\left( \varPhi _{n,1}\left( \lambda \right) ,\varPsi _{n,2}\left( \lambda \right) \right) =x_nWr\left( M_{n,1}\left( \lambda \right) ,N_{n,2}\left( \lambda \right) \right) ,
\end{equation}
\begin{equation}\tag{2.10b}
\bar{a}\left( \lambda \right) =x_nWr\left( \varPsi _{n,1}\left( \lambda \right) ,\varPhi _{n,2}\left( \lambda \right) \right) =x_nWr\left( N_{n,1}\left( \lambda \right) ,M_{n,2}\left( \lambda \right) \right) ,
\end{equation}
\begin{equation}\tag{2.10c}
b\left( \lambda \right) =\lambda ^{2n}x_nWr\left( N_{n,1}\left( \lambda \right) ,M_{n,1}\left( \lambda \right) \right) ,\ \ \bar{b}\left( \lambda \right) =\lambda ^{-2n}x_nWr\left( M_{n,2}\left( \lambda \right) ,N_{n,2}\left( \lambda \right) \right) ,
\end{equation}
with $x_n=\Delta_n|_{c_0=0}$(defined below Eqs.~(3.7)),
which provides a direct relationship between the scattering coefficients and the Jost eigenfunctions. The two symmetries in Proposition 2.1 ensure that the discrete eigenvalues exist in groups of $
\left\{ \lambda _j,\ -\lambda _j,\ \lambda _{j}^{-1},\ -\lambda _{j}^{-1} \right\}$ $\left( j=1,...,J \right).$

\vspace{5mm}\noindent\textbf{2.2 Inverse problem }

In order to construct the inverse problem, it is necessary to introduce a new matrix function $
\mu _n\left( \lambda \right) =M_n\left( \lambda \right)  \left(
                      \begin{array}{cc}
                        \frac{1}{a\left( \lambda \right)} & 0 \\
                        0 & \frac{1}{\bar{a}\left( \lambda \right)} \\
                      \end{array}
                    \right),
$ its two columns are respectively meromorphic in $D_{out}$ and $D_{in}$ and have finite poles. To remove the dependence of modified eigenfunctions on $u_k$ at the boundary, we modify them again as
\begin{equation}\tag{2.11a}
\tilde{N}_n\left( \lambda \right) =\left( \begin{matrix}
	1&		0\\
	0&		x_n\\
\end{matrix} \right) N_n\left( \lambda \right) =\left( \begin{matrix}
	x_{n}^{-1}&		-\lambda ^{-1}x_{n}^{-1}\frac{2bu_n+a}{\sqrt{\sigma \left( a^2+4b \right)}}\\
	-\lambda \sigma \frac{2bu_n+a}{\sqrt{\sigma \left( a^2+4b \right)}}&		1\\
\end{matrix} \right) +\left( \begin{matrix}
	O\left( \lambda ^2 \right)&		O\left( \lambda ^{-2} \right)\\
\end{matrix} \right) ,
\end{equation}
\begin{equation}\tag{2.11b}
\tilde{\mu}_n\left( \lambda \right) =\left( \begin{matrix}
	1&		0\\
	0&		x_n\\
\end{matrix} \right) \mu _n\left( \lambda \right) =\left( \begin{matrix}
	1&		\lambda \frac{2bu_{n-1}+a}{\sqrt{\sigma \left( a^2+4b \right)}}\\
	\lambda ^{-1}x_n\sigma \frac{2bu_{n-1}+a}{\sqrt{\sigma \left( a^2+4b \right)}}&		x_n\\
\end{matrix} \right) +\left( \begin{matrix}
	O\left( \lambda ^{-2} \right)&		O\left( \lambda ^2 \right)\\
\end{matrix} \right).
\end{equation}
Recalling the first property in Proposition 2.1, we obtain the jump relationship between two meromorphic matrix functions on $|\lambda|=1$
\begin{equation}\tag{2.12}
\left( \begin{matrix}
	\tilde{\mu}_{n,1}\left( \lambda \right)&		\tilde{N}_{n,2}\left( \lambda \right)\\
\end{matrix} \right) =\left( \begin{matrix}
	\tilde{N}_{n,1}\left( \lambda \right)&		\tilde{\mu}_{n,2}\left( \lambda \right)\\
\end{matrix} \right) \left( I+\left( \begin{matrix}
	-\rho \left( \lambda \right) \bar{\rho}\left( \lambda \right)&		-\lambda ^{2n}\bar{\rho}\left( \lambda \right)\\
	\lambda ^{-2n}\rho \left( \lambda \right)&		0\\
\end{matrix} \right) \right)
\end{equation}
in which $
\rho \left( \lambda \right) =\frac{b\left( \lambda \right)}{a\left( \lambda \right)}
$, $
\bar{\rho}\left( \lambda \right) =\frac{\bar{b}\left( \lambda \right)}{\bar{a}\left( \lambda \right)}
$ are reflection coefficients,
and
$$
\left( \begin{matrix}
	\tilde{\mu}_{n,1}\left( \lambda \right)&		\tilde{N}_{n,2}\left( \lambda \right)\\
\end{matrix} \right) \rightarrow I\ \ \ \ \text{as}\ \ \lambda \rightarrow \pm \infty ,
$$
$$
\left( \begin{matrix}
	\tilde{N}_{n,1}\left( \lambda \right)&		\tilde{\mu}_{n,2}\left( \lambda \right)\\
\end{matrix} \right) \rightarrow I\,\,\,\,\,\,\,\,\text{as}\,\,\,\,\lambda \rightarrow 0.
$$
In the below sections, we will classify and discuss the solutions of Eq.~(1.2) based on the different situations of poles.

\vspace{5mm}\noindent\textbf{2.3 Case of one-order poles }

In this section, we study the case as $a(\lambda)$ and $\bar{a}(\lambda)$ only have simple zeros resulting
that there exist two constants $b_j$ and $\bar{b}_j$ such that
$$
\varPhi _{n,1}\left( \lambda _j \right) =b_j\varPsi _{n,2}\left( \lambda _j \right) ,\ \ \varPhi _{n,2}\left( \bar{\lambda}_j \right) =\bar{b}_j\varPsi _{n,1}\left( \bar{\lambda}_j \right)
$$ for $a(\lambda_j)=0$ and $\bar{a}(\bar{\lambda}_j)=0$. Then we have the residues conditions for $\mu_n$
\begin{equation}\tag{2.13a}
\text{Res}\left( \mu _{n,1};\lambda _j \right) =\frac{M_{n,1}\left( \lambda _j \right)}{a'\left( \lambda _j \right)}=\frac{b_j\lambda _{j}^{-2n}N_{n,2}\left( \lambda _j \right)}{a'\left( \lambda _j \right)}=C_j\lambda _{j}^{-2n}N_{n,2}\left( \lambda _j \right) ,
\end{equation}
\begin{equation}\tag{2.13b}
\text{Res}\left( \mu _{n,2};\bar{\lambda}_j \right) =\frac{M_{n,2}\left( \bar{\lambda}_j \right)}{\bar{a}'\left( \bar{\lambda}_j \right)}=\frac{\bar{b}_j\bar{\lambda}_{j}^{2n}N_{n,1}\left( \bar{\lambda}_j \right)}{\bar{a}'\left( \bar{\lambda}_j \right)}=\bar{C}_j\bar{\lambda}_{j}^{2n}N_{n,1}\left( \bar{\lambda}_j \right) .
\end{equation}
According to the inverse problem we have constructed in the previous section, and considering the analytic properties of the modified eigenfunctions and scattering coefficients, we have
\begin{equation}\tag{2.14a}
\tilde{\mu}_{n,21}\left( \lambda \right) =\sum_{j=1}^J{\bar{C}_j\bar{\lambda}_{j}^{2n}\tilde{N}_{n,11}\left( \bar{\lambda}_j \right) \left( \frac{1}{\lambda -\bar{\lambda}_j}+\frac{1}{\lambda +\bar{\lambda}_j} \right)}+\frac{1}{2\pi i}\oint_{\left| \omega \right|=1}{\frac{\omega ^{2n}\bar{\rho}\left( \omega \right) \tilde{N}_{n,11}\left( \omega \right)}{\omega -\lambda}d\omega},
\end{equation}
\begin{equation}\tag{2.14b}
\tilde{N}_{n,11}\left( \lambda \right) =1+\sum_{j=1}^J{C_j\lambda _{j}^{-2n}\tilde{N}_{n,21}\left( \lambda _j \right) \left( \frac{1}{\lambda -\lambda _j}-\frac{1}{\lambda +\lambda _j} \right)}-\frac{1}{2\pi i}\oint_{\left| \omega \right|=1}{\frac{\omega ^{-2n}\rho \left( \omega \right) \tilde{N}_{n,21}\left( \omega \right)}{\omega -\lambda}d\omega},
\end{equation}
\begin{equation}\tag{2.14c}
\tilde{N}_{n,21}\left( \lambda \right) =\sum_{j=1}^J{\bar{C}_j\bar{\lambda}_{j}^{2n}\tilde{N}_{n,11}\left( \bar{\lambda}_j \right) \left( \frac{1}{\lambda -\bar{\lambda}_j}+\frac{1}{\lambda +\bar{\lambda}_j} \right)}+\frac{1}{2\pi i}\oint_{\left| \omega \right|=1}{\frac{\omega ^{2n}\bar{\rho}\left( \omega \right) \tilde{N}_{n,11}\left( \omega \right)}{\omega -\lambda}d\omega},
\end{equation}the second symmetry is used here. Comparing Eq.~(2.14a) with Eq.~(2.11b), we successfully reconstruct the potential function of Eq.~(1.2) as follows
\begin{equation}\tag{2.15}
u_{n-1}=\frac{\sqrt{\sigma \left( a^2+4b \right)}}{2b}\left( -2\sum_{j=1}^J{\bar{C}_j\bar{\lambda}_{j}^{2\left( n-1 \right)}\tilde{N}_{n,11}\left( \bar{\lambda}_j \right)}+\frac{1}{2\pi i}\oint_{\left| \omega \right|=1}{\omega ^{2\left( n-1 \right)}\bar{\rho}\left( \omega \right) \tilde{N}_{n,11}\left( \omega \right) d\omega} \right) -\frac{a}{2b}.
\end{equation}
If reflection coefficients equal to zero, the unknowns in Eq.~(2.15) include the value of $\tilde{N}_{n,11}\left( \bar{\lambda}_j \right)$ and the dependence of $\bar{C}_j$ on time, which are the problems we need to solve next. From Eqs.~(2.14b-c), we get a system of linear equations
$$
\left\{ \begin{array}{l}
	\tilde{N}_{n,11}\left( \bar{\lambda}_i \right) =1+\sum_{j=1}^J{C_j\lambda _{j}^{-2n}\tilde{N}_{n,21}\left( \lambda _j \right) \left( \frac{1}{\bar{\lambda}_i-\lambda _j}-\frac{1}{\bar{\lambda}_i+\lambda _j} \right)}-\frac{1}{2\pi i}\oint_{\left| \omega \right|=1}{\frac{\omega ^{-2n}\rho \left( \omega \right) \tilde{N}_{n,21}\left( \omega \right)}{\omega -\bar{\lambda}_i}d\omega},\\
	\tilde{N}_{n,21}\left( \lambda _i \right) =\sum_{j=1}^J{\bar{C}_j\bar{\lambda}_{j}^{2n}\tilde{N}_{n,11}\left( \bar{\lambda}_j \right) \left( \frac{1}{\lambda _i-\bar{\lambda}_j}+\frac{1}{\lambda _i+\bar{\lambda}_j} \right)}+\frac{1}{2\pi i}\oint_{\left| \omega \right|=1}{\frac{\omega ^{2n}\bar{\rho}\left( \omega \right) \tilde{N}_{n,11}\left( \omega \right)}{\omega -\lambda _i}d\omega}.\\
\end{array} \right.
$$Assuming there is only one set of discrete eigenvalues and no reflection, then
\begin{equation}\tag{2.16}
\tilde{N}_{n,11}\left( \bar{\lambda}_1 \right) =\frac{\det \left( A_1 \right)}{\det \left( A \right)},
\end{equation}
with
$$
A_1=\left( \begin{matrix}
	1&		-C_1\lambda _{1}^{-2n}\left( \frac{1}{\bar{\lambda}_1-\lambda _1}-\frac{1}{\bar{\lambda}_1+\lambda _1} \right)\\
	0&		1\\
\end{matrix} \right) ,\ \ A=\left( \begin{matrix}
	1&		-C_1\lambda _{1}^{-2n}\left( \frac{1}{\bar{\lambda}_1-\lambda _1}-\frac{1}{\bar{\lambda}_1+\lambda _1} \right)\\
	-\bar{C}_1\bar{\lambda}_{1}^{2n}\left( \frac{1}{\lambda _1-\bar{\lambda}_1}+\frac{1}{\lambda _1+\bar{\lambda}_1} \right)&		1\\
\end{matrix} \right) .
$$
Considering Proposition 2.1 and the definitions norming constants, we have \begin{equation}\tag{2.17}
\bar{b}_j=\pm b_{j},\
\bar{a}'\left( \bar{\lambda}_j \right) =-\lambda _{j}^{2}a'\left( \lambda _j \right),\
\bar{C}_j=\pm \lambda _{j}^{-2}C_j.
\end{equation}

Now we take account of time evolution, one can redefine two time-dependent eigenfunctions
\begin{equation}\tag{2.18}
\hat{M}_n\left( \lambda ,t \right) =M_n\left( \lambda ,t \right) \left( \begin{matrix}
	e^{\omega _1t}&		0\\
	0&		e^{\omega _2t}\\
\end{matrix} \right) ,\ \ \hat{N}_n\left( \lambda ,t \right) =N_n\left( \lambda ,t \right) \left( \begin{matrix}
	e^{\omega _1t}&		0\\
	0&		e^{\omega _2t}\\
\end{matrix} \right)
\end{equation}
recalling that $\frac{d\psi _n}{dt}=\left( \begin{matrix}
	\omega _1&		0\\
	0&		\omega _2\\
\end{matrix} \right) \psi _n$ as $
n\rightarrow \pm \infty
$. The relationship between them can be obtained from Eqs.~(2.2) and Proposition 2.1 as
\begin{equation}\tag{2.19}
\hat{M}_n\left( \lambda ,t \right) =\hat{N}_n\left( \lambda ,t \right) \left( \begin{matrix}
	a\left( \lambda ,t \right)&		\lambda ^{2n}e^{\omega _2t}\bar{b}\left( \lambda ,t \right)\\
	\lambda ^{-2n}e^{\omega _1t}b\left( \lambda ,t \right)&		\bar{a}\left( \lambda ,t \right)\\
\end{matrix} \right).
\end{equation}
Through taking the direct derivative of $t$ in the above equation and comparing it with the result of satisfying Lax pair at the boundary, we deduce the dependence of scattering coefficients and norm constant on time as
\begin{equation}\tag{2.20a}
a\left( \lambda ,t \right) =a\left( \lambda ,0 \right) ,\ \ \bar{a}\left( \lambda ,t \right) =\bar{a}\left( \lambda ,0 \right) ,
\end{equation}
\begin{equation}\tag{2.20b}
b\left( \lambda ,t \right) =b\left( \lambda ,0 \right) e^{\left( \omega _2\left( \lambda \right) -\omega _1\left( \lambda \right) \right) t},\ \ \bar{b}\left( \lambda ,t \right) =\bar{b}\left( \lambda ,0 \right) e^{\left( \omega _1\left( \lambda \right) -\omega _2\left( \lambda \right) \right) t},
\end{equation}
\begin{equation}\tag{2.20c}
b_j\left( t \right) =b_j\left( 0 \right) e^{\left( \omega _2\left( \lambda _j \right) -\omega _1\left( \lambda _j \right) \right) t},\ \ C_j\left( t \right) =C_j\left( 0 \right) e^{\left( \omega _2\left( \lambda _j \right) -\omega _1\left( \lambda _j \right) \right) t}.
\end{equation} Note that if $\sigma =1$, there is singularity in the reflectless potential, so we focus on the case of $\sigma =-1$ here. Combining Eqs.~(2.15)-(2.17) and (2.20c), we obtain the one-soliton solution after simple calculations and simplifications
\begin{equation}\tag{2.21}
u_n=-\frac{a}{2b}+\text{sgn}(C_1\left( 0 \right))\frac{\sqrt{-\left( a^2+4b \right)} \left( \lambda _{1}^{-2}-\lambda _{1}^{2} \right)}{4b}\text{sech}\left[ p_1 +\text{log}\left( \frac{\lambda _{1}^{4}-1}{2|C_1\left( 0 \right)|} \right) \right] ,
\end{equation}
$$
p_1=2n\text{log}\left( \lambda _1 \right) +t\left( \omega _1\left( \lambda _1 \right) -\omega _2\left( \lambda _1 \right) \right),
$$
propagating with velocity $V_s=\frac{\omega _2\left( \lambda _1 \right) -\omega _1\left( \lambda _1 \right)}{2Ln\left( \lambda _1 \right)}$ on the background $-\frac{a}{2b}$. Owing to $a^2+4b<0$ must be satisfied, we can obtain bright soliton as $C_1(0)>0$ and dark soliton when $C_1(0)<0$ (see Fig.~1).

\begin{center}
\includegraphics[scale=0.5]{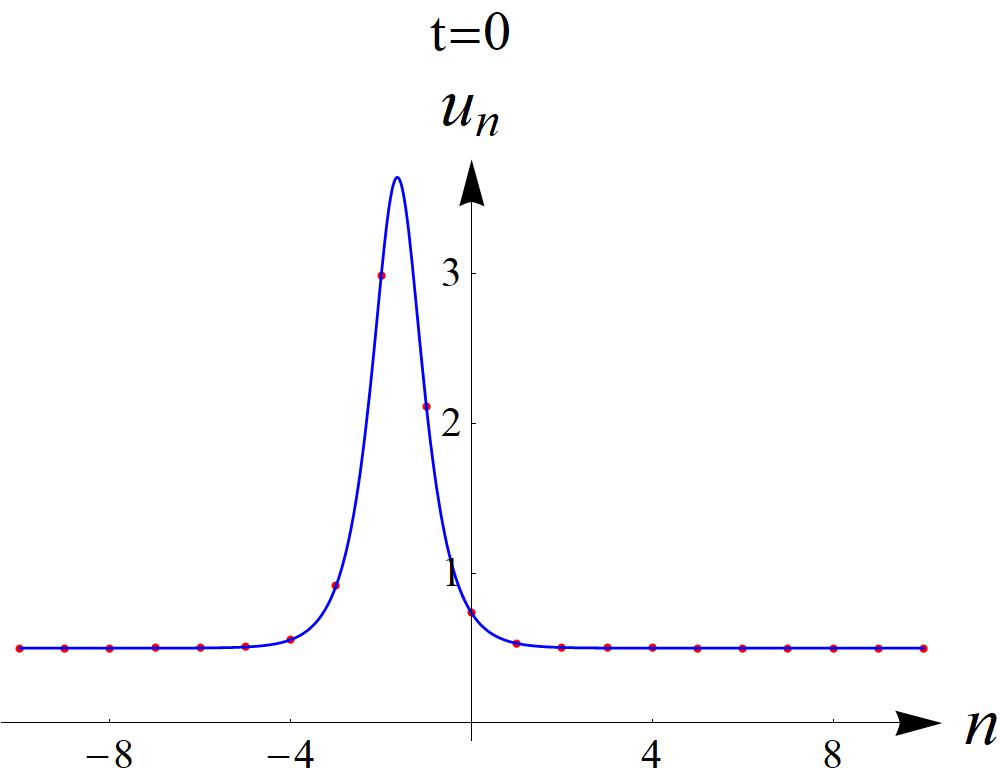}\hspace{2.3cm}
\includegraphics[scale=0.5]{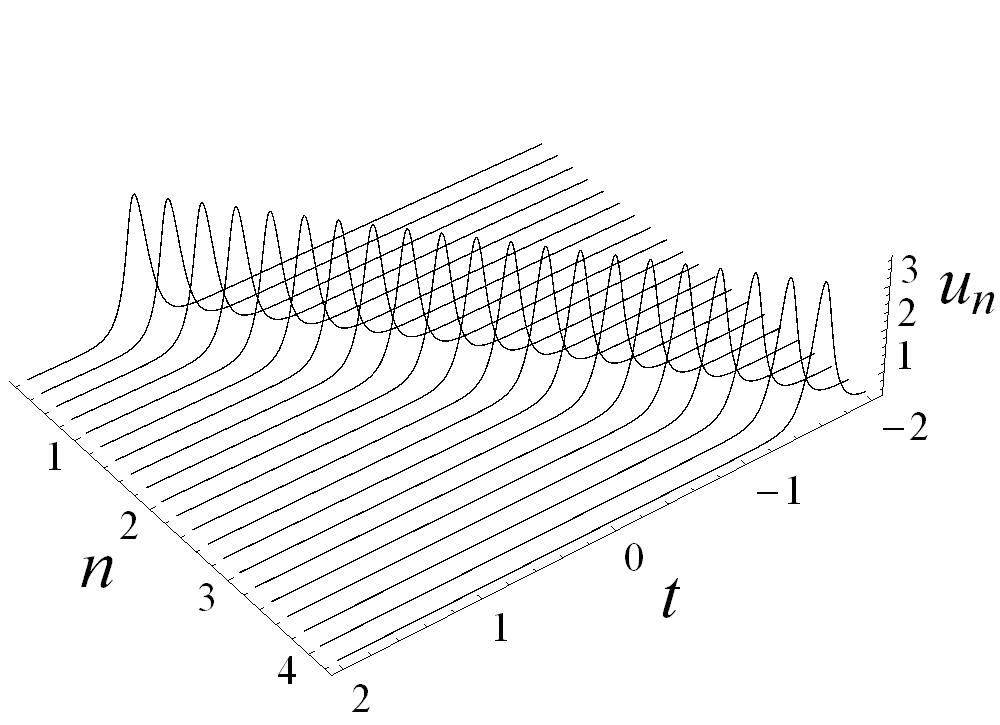}

\vspace{-0.2cm}{\footnotesize\hspace{0.1cm}(a)\hspace{7.5cm}(b)}\\
\includegraphics[scale=0.5]{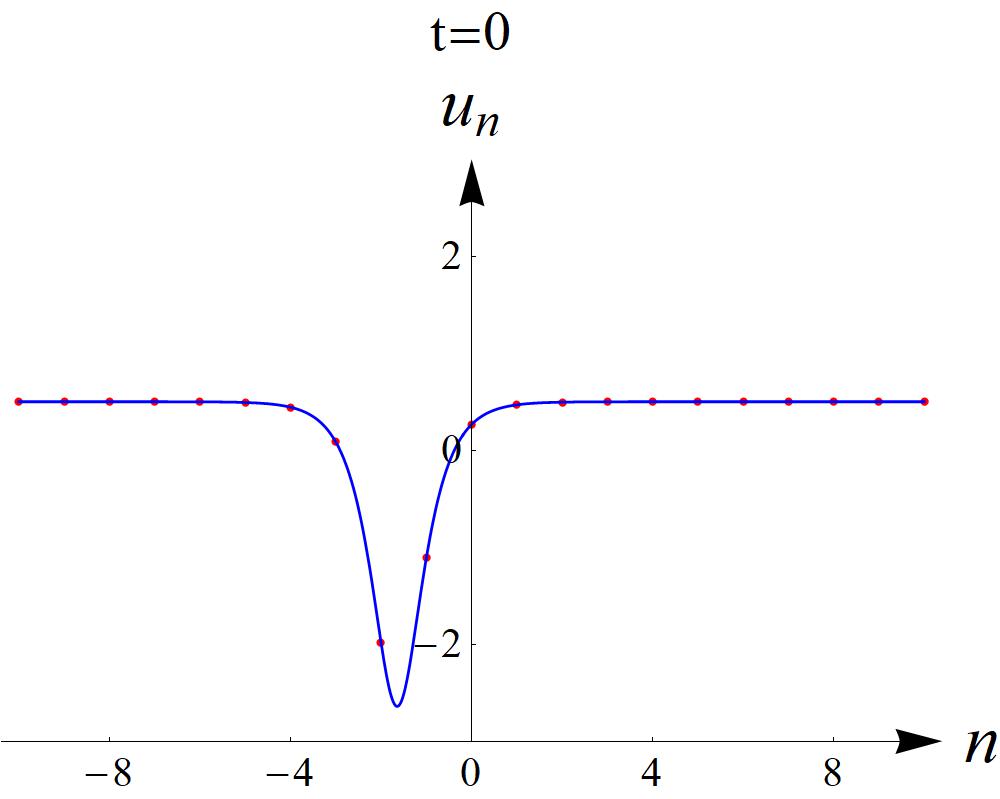}\hspace{2.3cm}
\includegraphics[scale=0.5]{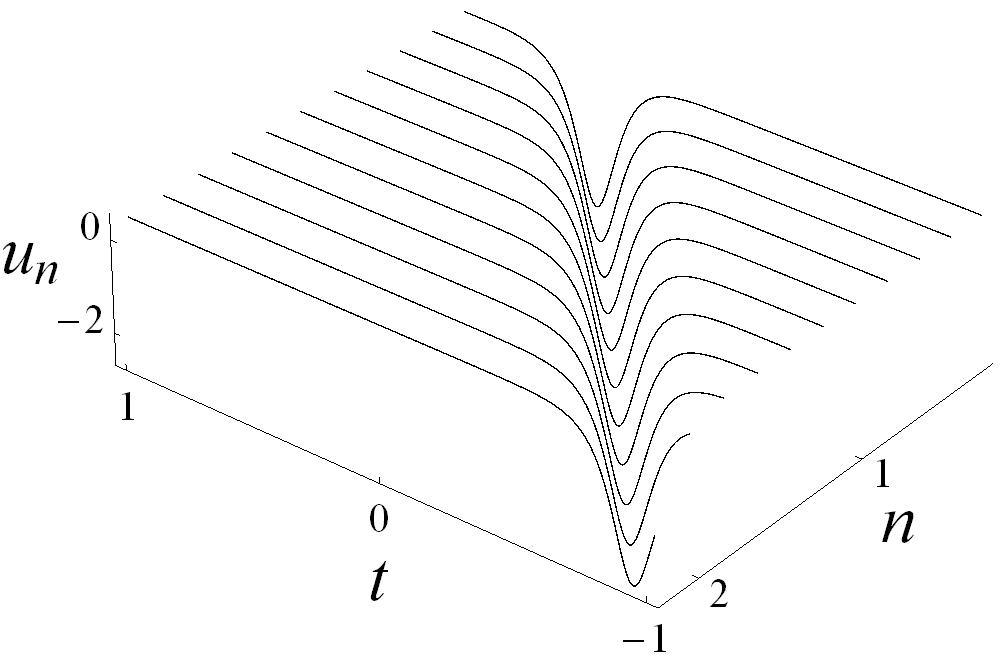}

\vspace{-0.2cm}{\footnotesize\hspace{0.0cm}(c)\hspace{7.5cm}(d)}\\
\flushleft{\footnotesize
\textbf{Fig.~$1$.} Solutions corresponding to Eq.~(2.21): (a),(b) the bright soliton on a positive background with $a=1$, $b=-1$, $\lambda_1=e$, $C_1(0)=1$; (c),(d) the dark soliton on a positive background with $a=1$, $b=-1$, $\lambda_1=e$, $C_1(0)=-1$. The soliton solution on a negative background can be obtained by taking a negative value of $a$.}
\end{center}

When there are two sets of discrete eigenvalues, Eq.~(1.2) admits the collision between two solitons including the head-on collision and the overtaking collision in term of
$$
u_n=-\frac{a}{2b}+\frac{\sqrt{-\left( a^2+4b \right)}}{2b}\frac{f}{g}
$$
with
\begin{align}\notag
f=&-2\left( \lambda _{1}^{2}\lambda _{2}^{2}-1 \right) ^2\left( 4z_1\lambda _{1}^{2}\left( \lambda _{1}^{2}-\lambda _{2}^{2} \right) ^2\left( \lambda _{2}^{4}-1 \right) ^2+z_{1}^{-1}\lambda _{1}^{2}\left( \lambda _{1}^{4}-1 \right) ^2\left( \lambda _{1}^{2}\lambda _{2}^{2}-1 \right) ^2\left( \lambda _{2}^{4}-1 \right) ^2\right.\\\notag
&\ \ \ \ \ \ \ \ \ \ \ \left.+\left( \lambda _{1}^{4}-1 \right) ^2\lambda _{2}^{2}\left( 4z_2\left( \lambda _{1}^{2}-\lambda _{2}^{2} \right) ^2+z_{2}^{-1}\left( \lambda _{1}^{2}\lambda _{2}^{2}-1 \right) ^2\left( \lambda _{2}^{4}-1 \right) ^2 \right) \right),\\\notag
g=&\lambda _{1}^{2}\lambda _{2}^{2}\left( 8\left( \lambda _{1}^{4}-1 \right) ^2\left( \lambda _{1}^{2}\lambda _{2}^{2}-1 \right) ^2\left( \lambda _{2}^{4}-1 \right) ^2+z_{1}^{-1}\left( \lambda _{1}^{4}-1 \right) ^2\left( \lambda _{1}^{2}\lambda _{2}^{2}-1 \right) ^4\left( 4z_2+z_{2}^{-1}\left( \lambda _{2}^{4}-1 \right) ^2 \right) \right.\\\notag
&\ \ \ \ \ \ \ \ \ \ \ \left.+4z_1\left( 4z_2\left( \lambda _{1}^{2}-\lambda _{2}^{2} \right) ^4+z_{2}^{-1}\left( \lambda _{1}^{2}\lambda _{2}^{2}-1 \right) ^4\left( \lambda _{2}^{4}-1 \right) ^2 \right) \right),\notag
\end{align}
$$
z_{i}^{-1}=C_1\left( 0 \right) ^{-1}e^{p_i},\ \ p_i=2n\text{log}\left[ \lambda _i \right] +\left( \omega _1\left( \lambda _i \right) -\omega _2\left( \lambda _i \right) \right) t,\ \ i=1,2.
$$
To understand the dynamic properties of the two-solitons, we analyze the asymptotic behaviors of them $$
\frac{f}{g}\sim \frac{1}{2}\text{sgn}\left( C_j\left( 0 \right) \right) \left( \lambda _{j}^{-2}-\lambda _{j}^{2} \right) \text{sech}\left( p_j+\xi _{j\pm} \right) ,\ \ j=2-i,\ \ \ p_i\sim \pm \infty ,
$$
$$
\xi _{j+}=\text{log}\left( \frac{\left( \lambda _{j}^{4}-1 \right)}{2\left| C_j\left( 0 \right) \right|} \right) ,\ \ \ \xi _{j-}=\text{log}\left( \frac{\left( \lambda _{j}^{4}-1 \right) \left( \lambda _{1}^{2}\lambda _{2}^{2}-1 \right) ^2}{2\left| C_j\left( 0 \right) \right|\left( \lambda _{1}^{2}-\lambda _{2}^{2} \right) ^2} \right).
$$
Now, the collision of two solitons in the one-order poles case has been proven to be elastic, namely, apart from the phase shift, the characteristics of the solitons, such as amplitude and velocity, remain unchanged before and after the collision and are identical to those of the one soliton solution mentioned earlier. This is not unexpected, as these characteristics of the solitons are completely determined by their corresponding discrete eigenvalues. Recalling that the propagation speed of solitons depends on $a$, $b$ and the discrete eigenvalues, the condition for its speed to be positive is equivalent to
$$
\lambda _{i}^{2}<\frac{-2\left( a^2-2b \right) -\sqrt{3}\sqrt{a^4-8a^2b}}{a^2+4b}.
$$
Furthermore, determining collision type is suggested by the requirement to focus on the sign structure of $V_{s1}V_{s2}$ for $b<0$, which is depicted in Fig.~2.
\begin{center}
\includegraphics[scale=0.8]{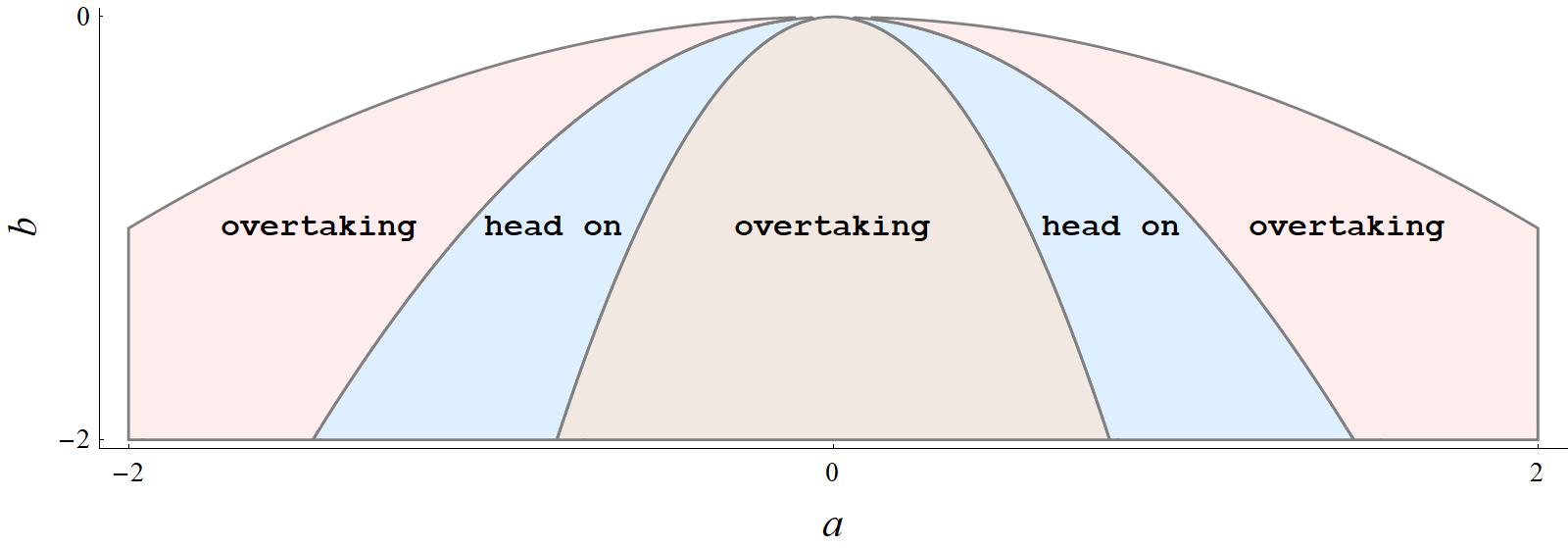}
\flushleft{\footnotesize
\textbf{Fig.~$2$.} Without loss of generality, we set $\lambda_2^2>\lambda_1^2>1$. The $ab$-plane is divided into five regions: the overtaking collision regions $Q>\lambda _{2}^{2}$ or $Q<\lambda _{1}^{2}$ and the head-on collision region $\lambda _{1}^{2}<Q<\lambda _{2}^{2}$ with $Q=\frac{-2\left( a^2-2b \right) -\sqrt{3}\sqrt{a^4-8a^2b}}{a^2+4b}$, $\lambda _{1}^{2}=2$, $\lambda _{2}^{2}=4$. The white region corresponds to $a^2+4b>0$ which is not within the scope of our consideration in this section.}
\end{center}

As analyzed in the previous section, the results of soliton-soliton interaction contain the two-bright-soliton solution, the two-dark-soliton solution and the bright-dark-soliton solution (see Figs.~3-6).
\begin{center}
\includegraphics[scale=0.43]{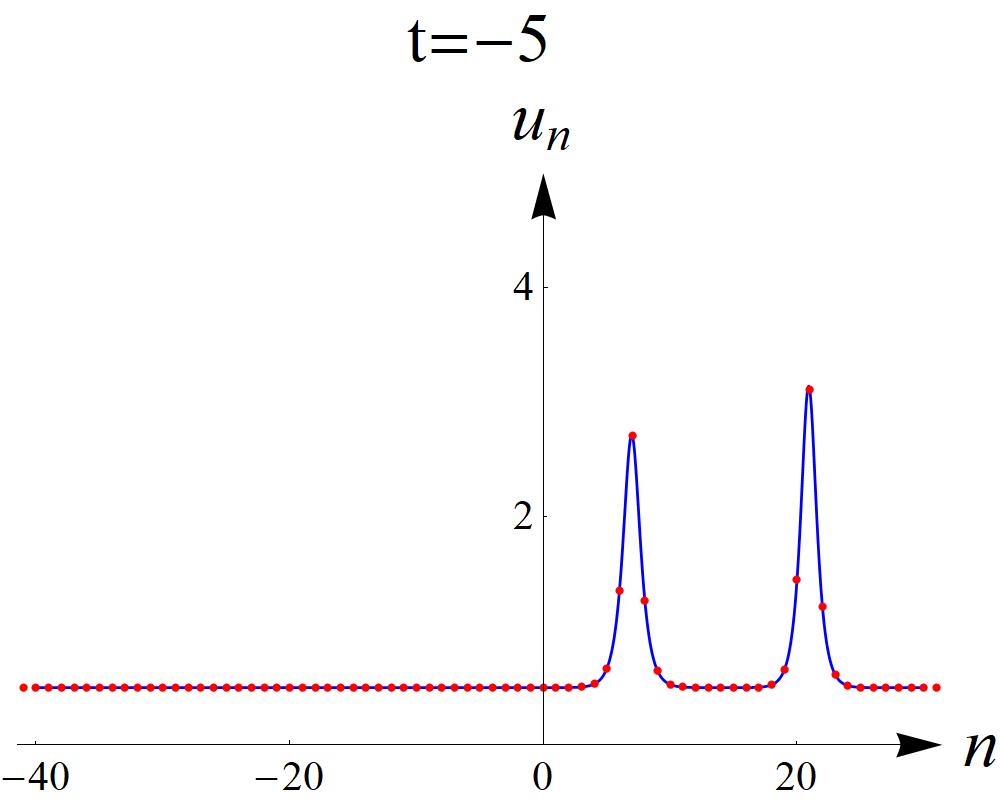}\hfill
\includegraphics[scale=0.43]{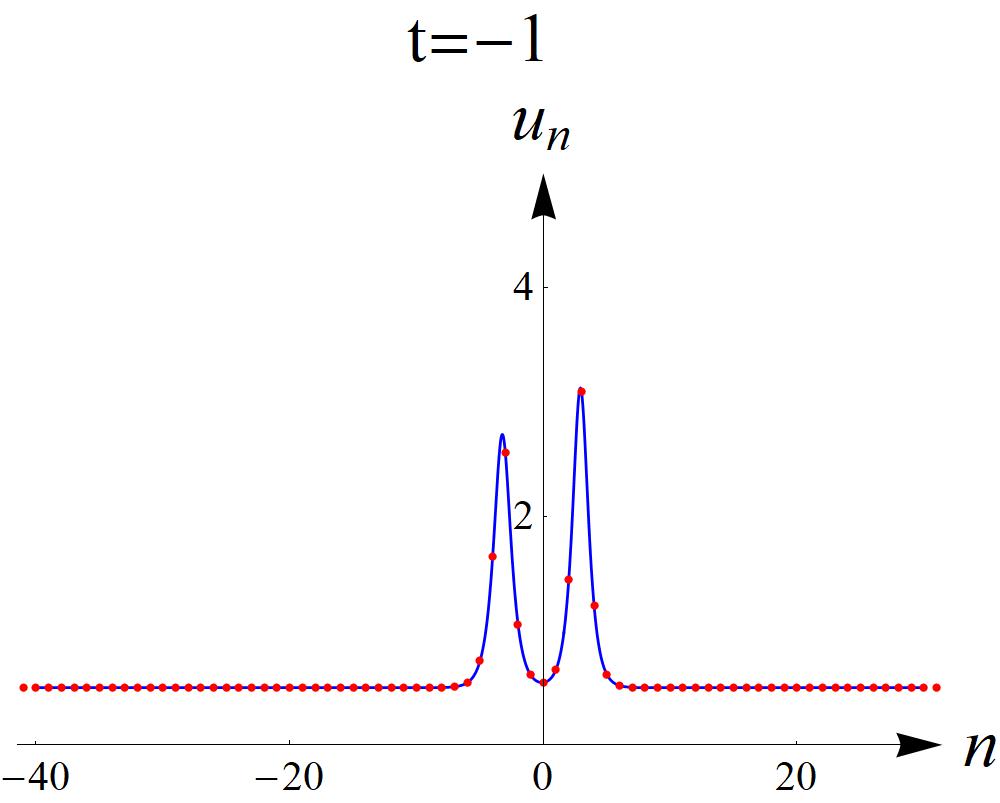}\hfill
\includegraphics[scale=0.43]{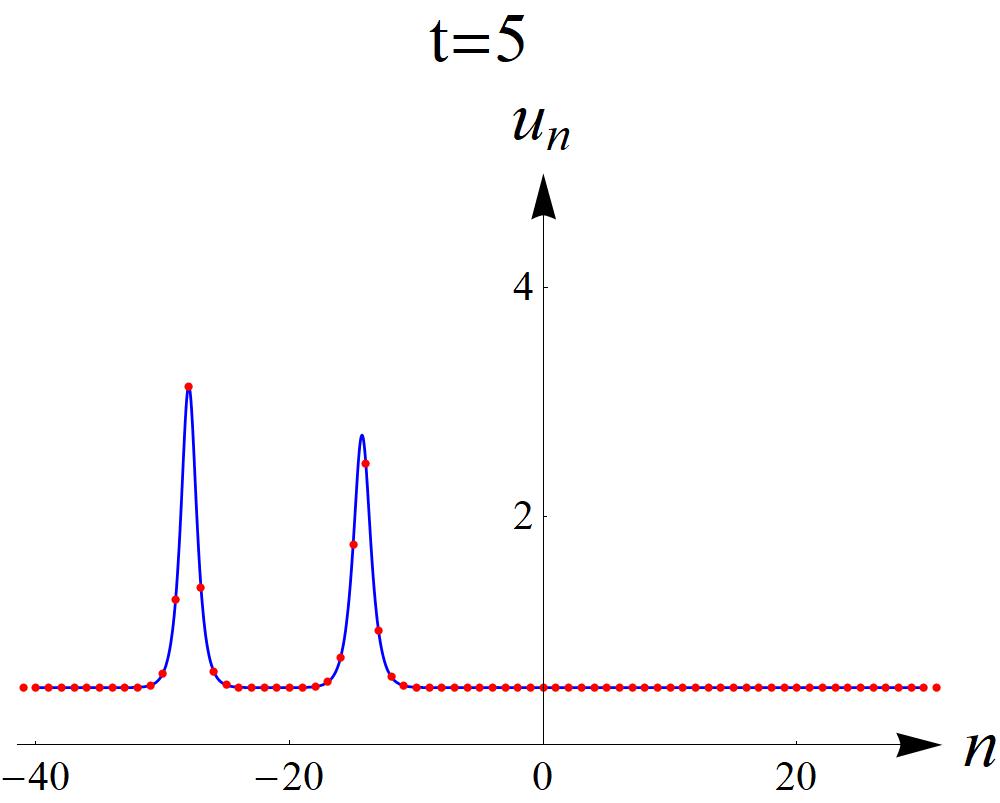}\hfill\\
\vspace{-0.2cm}{\footnotesize\hspace{0cm}(a)\hspace{5.9cm}(b)\hspace{5.9cm}(c)}
\flushleft{\footnotesize
\textbf{Figs.~$3$.} Profiles of evolutions of two overtaking collision bright solitons solution with $a=1$, $b=-1$, $\lambda_1=2.3$, $\lambda_2=2.5$, $C_1(0)=C_2(0)=1$.}
\end{center}
\begin{center}
\includegraphics[scale=0.43]{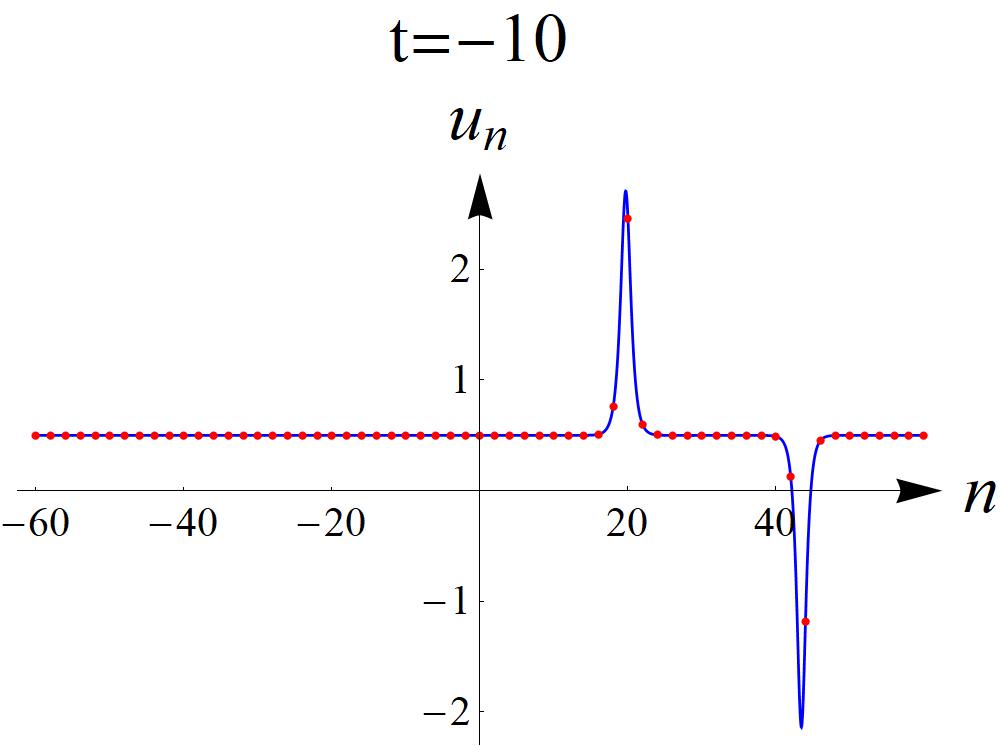}\hfill
\includegraphics[scale=0.43]{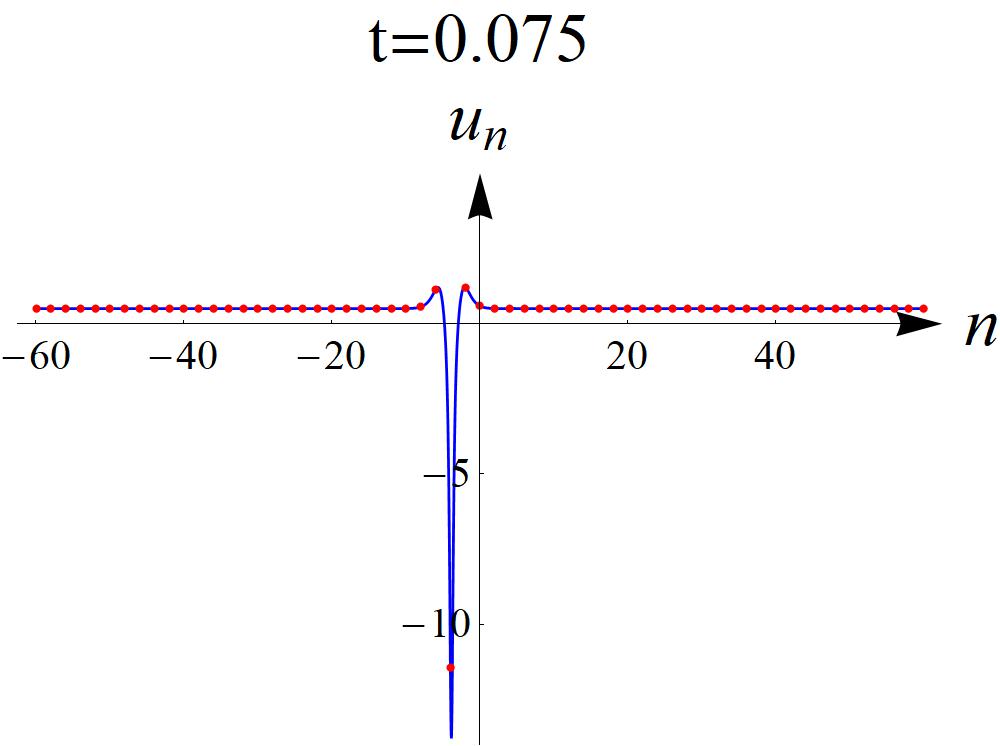}\hfill
\includegraphics[scale=0.43]{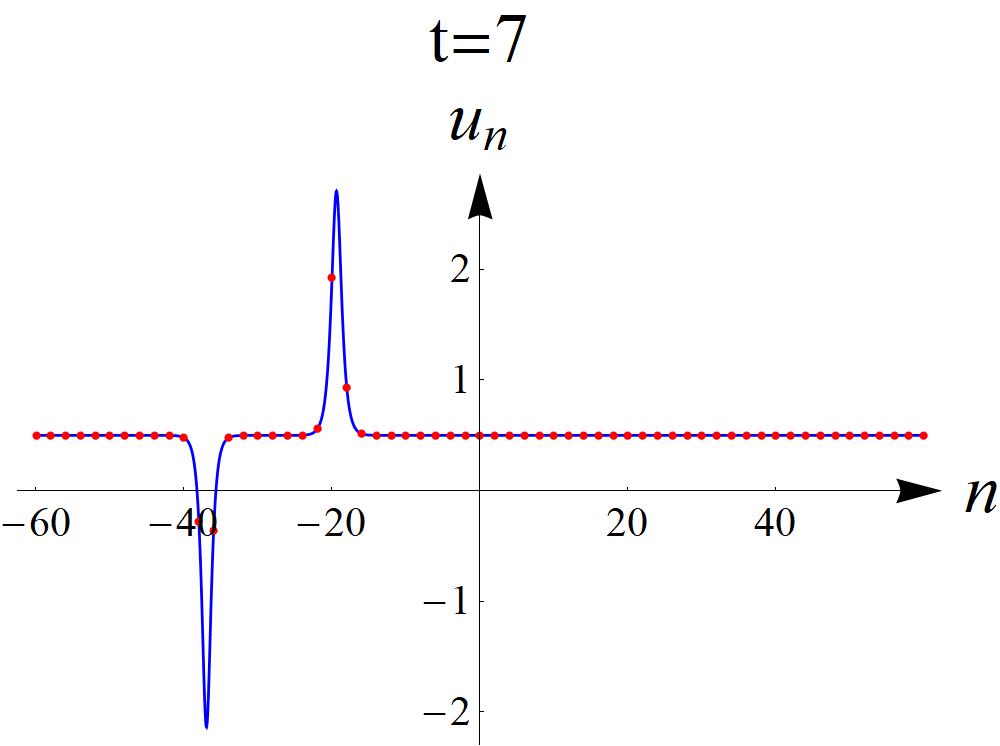}\hfill\\
\vspace{-0.2cm}{\footnotesize\hspace{0cm}(a)\hspace{5.9cm}(b)\hspace{5.9cm}(c)}
\flushleft{\footnotesize
\textbf{Figs.~$4$.} Profiles of evolutions of overtaking collision bright-dark solitons solution with $a=1$, $b=-1$, $\lambda_1=2.3$, $\lambda_2=2.5$, $C_1(0)=1$, $C_2(0)=-1$.}
\end{center}
\begin{center}
\includegraphics[scale=0.43]{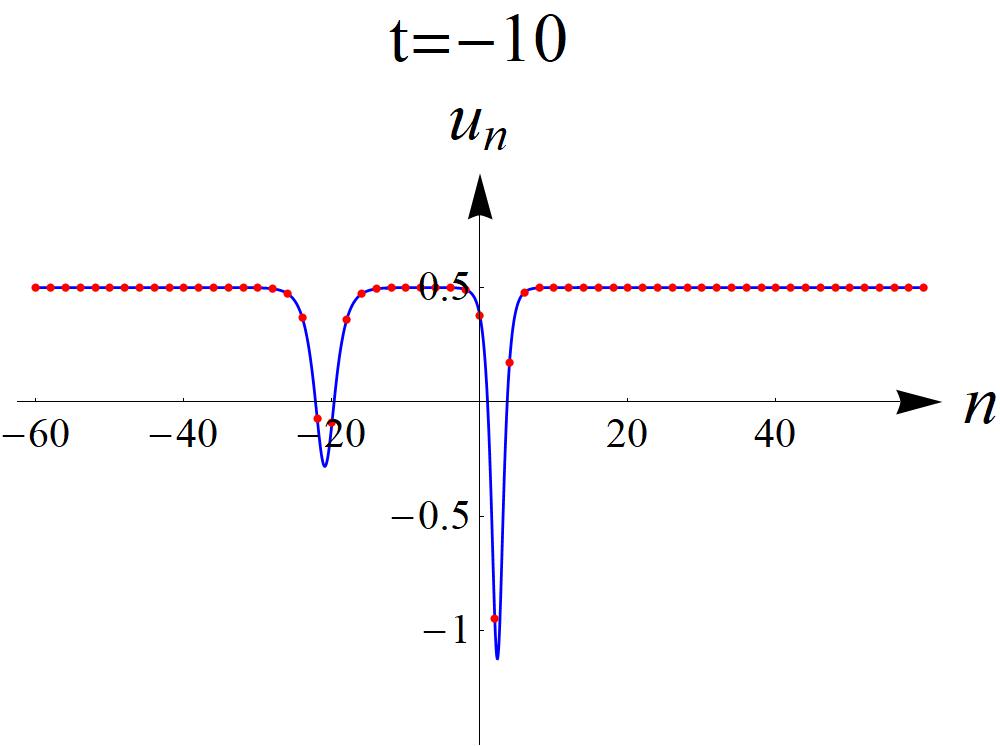}\hfill
\includegraphics[scale=0.43]{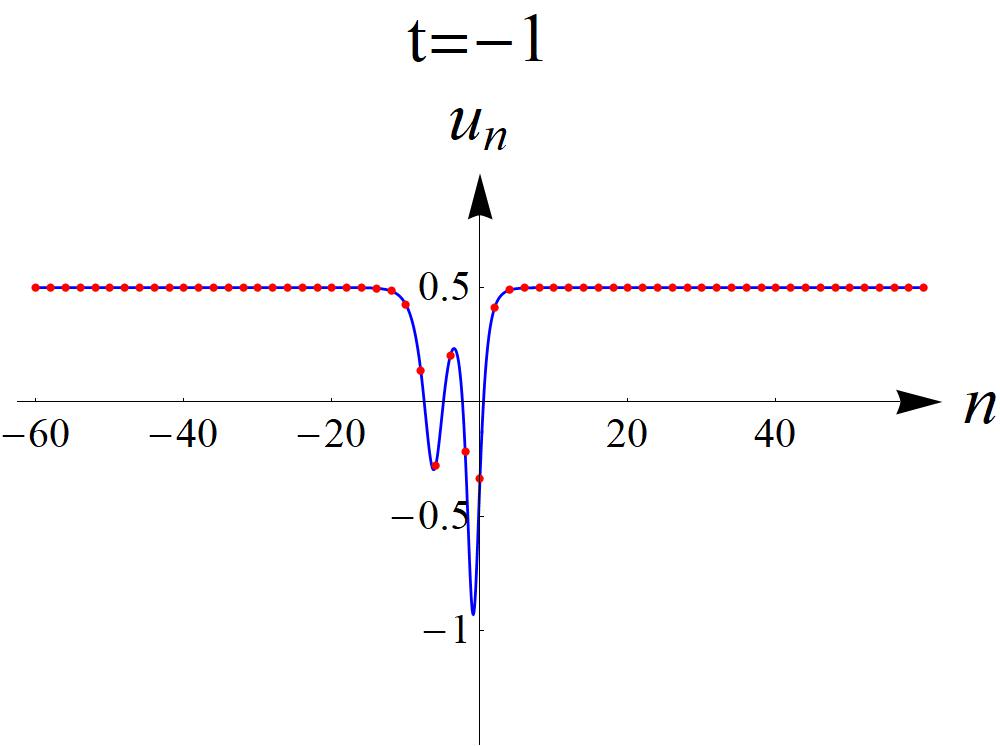}\hfill
\includegraphics[scale=0.43]{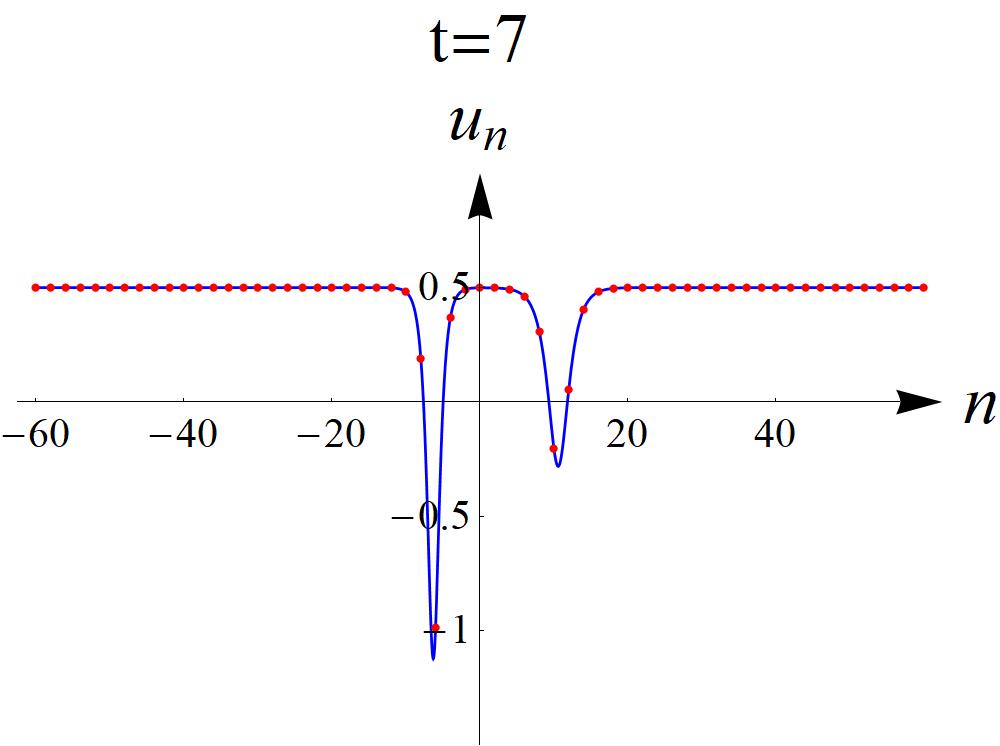}\hfill\\
\vspace{-0.2cm}{\footnotesize\hspace{0cm}(a)\hspace{5.9cm}(b)\hspace{5.9cm}(c)}
\flushleft{\footnotesize
\textbf{Figs.~$5$.} Profiles of evolutions of two head-on collision dark solitons solution with $a=1$, $b=-1$, $\lambda_1=2$, $\lambda_2=1.5$, $C_1(0)=C_2(0)=-1$.}
\end{center}
\newpage
\begin{center}
\includegraphics[scale=0.43]{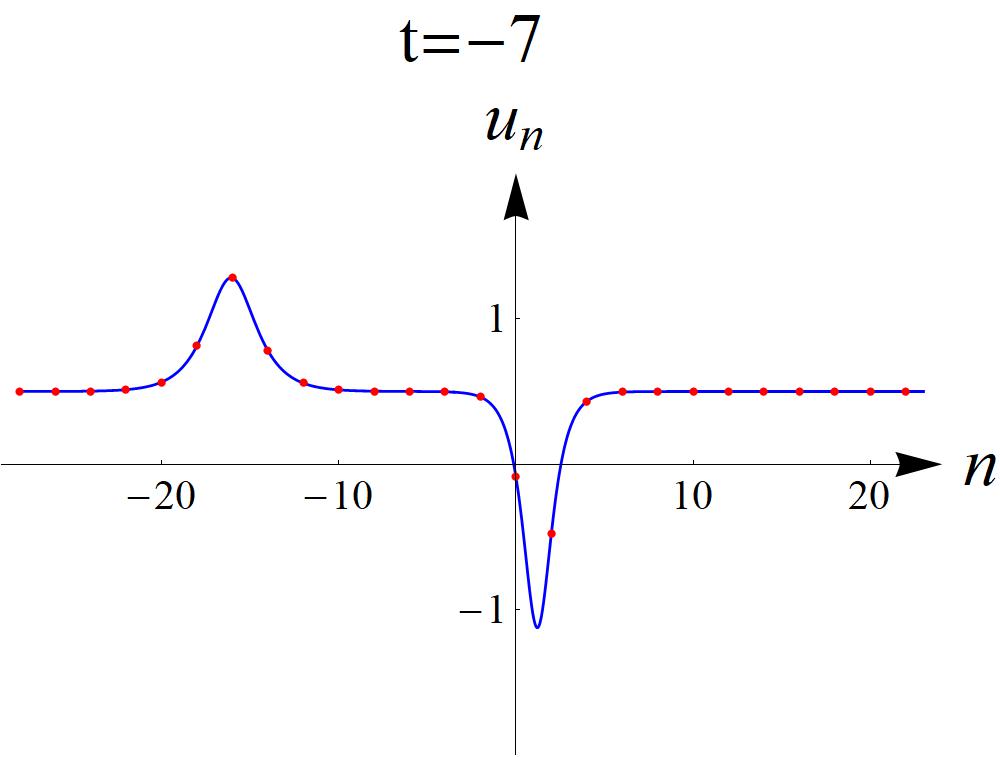}\hfill
\includegraphics[scale=0.43]{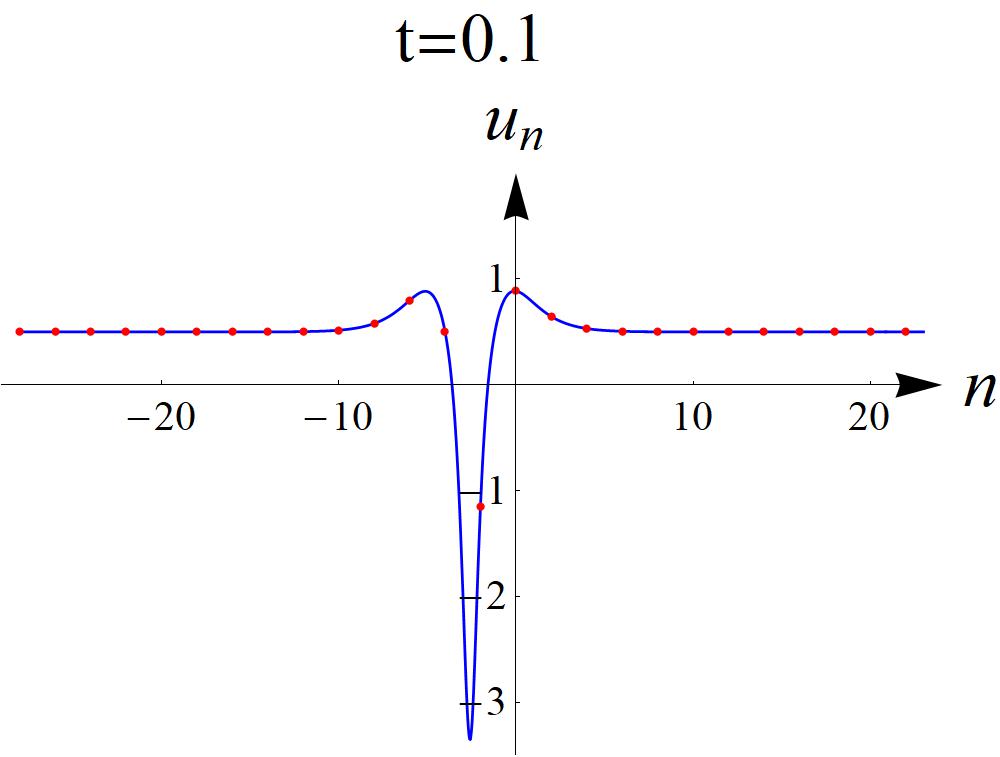}\hfill
\includegraphics[scale=0.43]{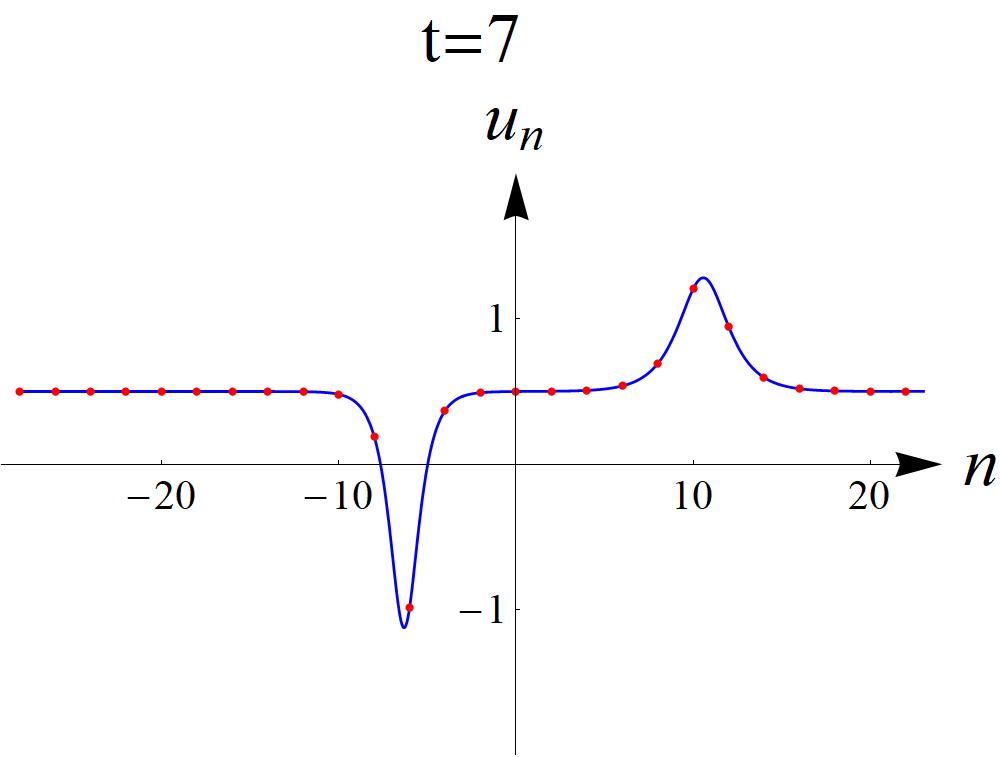}\hfill\\
\vspace{-0.2cm}{\footnotesize\hspace{0cm}(a)\hspace{5.9cm}(b)\hspace{5.9cm}(c)}
\flushleft{\footnotesize
\textbf{Figs.~$6$.} Profiles of evolutions of head-on collision bright-dark solitons solution with $a=1$, $b=-1$, $\lambda_1=1.5$, $\lambda_2=2$, $C_1(0)=1$, $C_2(0)=-1$.}
\end{center}

From the Figs.~3-6, it can be seen that two solitons with different polarities can generate the so-called rogue waves at the moment of collision, with amplitudes exceeding twice that of the original solitons. However, this phenomenon does not occur in the case of two solitons with the same polarity. Instead, the amplitude at the moment of collision is smaller than the maximum amplitude of the two solitons.

\vspace{5mm}\noindent\textbf{2.4 Case of two-order poles }

In this case, the zeros of scattering coefficients are double i.e, $a\left( \lambda _j \right) =a'\left( \lambda _j \right) =0,$ $\bar{a}\left( \bar{\lambda}_j \right) =\bar{a}'\left( \bar{\lambda}_j \right) =0.$ From Eq.~(2.10a),
$$
a'\left( \lambda _j \right) =x_nWr\left( \varPhi '_{n,1}\left( \lambda _j \right) -b_j\varPsi '_{n,2}\left( \lambda _j \right) ,\varPsi _{n,2}\left( \lambda _j \right) \right) =0,
$$
thus there exists a constant $d_j$ such that
$$
\varPhi '_{n,1}\left( \lambda _j \right) =b_j\varPsi '_{n,2}\left( \lambda _j \right) +d_j\varPsi _{n,2}\left( \lambda _j \right) .
$$
Then we can represent $M'_{n}\left( \lambda _j \right)$ in the form of $N_{n}\left( \lambda _j \right)$ as
\begin{equation}\tag{2.22a}
M'_{n,1}\left( \lambda _j \right) =\left( -2nb_j\lambda _{j}^{-2n-1}+d_j\lambda _{j}^{-2n} \right) N_{n,2}\left( \lambda _j \right) +b_j\lambda _{j}^{-2n}N'_{n,2}\left( \lambda _j \right) ,
\end{equation}
and
\begin{equation}\tag{2.22b}
M'_{n,2}\left( \bar{\lambda}_j \right) =\left( 2n\bar{b}_j\bar{\lambda}_{j}^{2n-1}+\bar{d}_j\bar{\lambda}_{j}^{2n} \right) N_{n,1}\left( \bar{\lambda}_j \right) +\bar{b}_j\bar{\lambda}_{j}^{2n}N'_{n,1}\left( \bar{\lambda}_j \right) ,
\end{equation}
similarly. The Laurent expansion of $\mu_n(\lambda)$ with respect to $\lambda_j$ or $\bar{\lambda}_j$ exists, naturally, to the power of -2, to which we must also consider the coefficients in addition to the residues at the poles. Recalling that $
\varPhi _{n,1}\left( \lambda _j \right) =b_j\varPsi _{n,2}\left( \lambda _j \right)
$ and Eq.~(2.22a), we have
\begin{equation}\tag{2.23a}
P_{-2}\left( \mu _{n,1},\lambda _j \right) =\frac{2M_{n,1}\left( \lambda _j \right)}{a''\left( \lambda _j \right)}=\frac{2b_j\lambda _{j}^{-2n}N_{n,2}\left( \lambda _j \right)}{a''\left( \lambda _j \right)}=F_j\lambda _{j}^{-2n}N_{n,2}\left( \lambda _j \right) ,
\end{equation}
\begin{align}\notag
P_{-1}\left( \mu _{n,1},\lambda _j \right) &=\frac{2M'_{n,1}\left( \lambda _j \right)}{a''\left( \lambda _j \right)}-\frac{2M_{n,1}\left( \lambda _j \right) a'''\left( \lambda _j \right)}{3\left( a''\left( \lambda _j \right) \right) ^2}\\ \notag
&=F_j\lambda _{j}^{-2n}N'_{n,2}\left( \lambda _j \right) +F_j\lambda _{j}^{-2n}\left( -2n\lambda _{j}^{-1}+D_j-G_j \right) N_{n,2}\left( \lambda _j \right) ,\tag{2.23b}
\end{align}
with $F_j=\frac{2b_j}{a''\left( \lambda _j \right)}$, $D_j=\frac{d_j}{b_j}$, $G_j=\frac{a'''\left( \lambda _j \right)}{3a''\left( \lambda _j \right)}$. For zeros $\bar{\lambda}_j$, we can obtain similar representations
\begin{equation}\tag{2.23c}
P_{-2}\left( \mu _{n,2},\bar{\lambda}_j \right) =\bar{F}_j\bar{\lambda}_{j}^{2n}N_{n,1}\left( \bar{\lambda}_j \right) ,
\end{equation}
\begin{equation}\tag{2.23d}
P_{-1}\left( \mu _{n,2},\bar{\lambda}_j \right) =\bar{F}_j\bar{\lambda}_{j}^{2n}N'_{n,1}\left( \bar{\lambda}_j \right) +\bar{F}_j\bar{\lambda}_{j}^{2n}\left( 2n\bar{\lambda}_{j}^{-1}+\bar{D}_j-\bar{G}_j \right) N_{n,1}\left( \bar{\lambda}_j \right) ,
\end{equation}
with $\bar{F}_j=\frac{2\bar{b}_j}{\bar{a}''\left( \bar{\lambda}_j \right)}$, $\bar{D}_j=\frac{\bar{d}_j}{\bar{b}_j}$, $\bar{G}_j=\frac{\bar{a}'''\left( \bar{\lambda}_j \right)}{3\bar{a}''\left( \bar{\lambda}_j \right)}$.

Here, we need to consider two types of symmetry properties for above new norming constants. According to Proposition 2.1, we can deduce the first symmetry property
\begin{equation}\tag{2.24}
d_{j}^{-}=d_{j}^{+},\ \ F_{j}^{-}=-F_{j}^{+},\ \ D_{j}^{-}=-D_{j}^{+},\ \ G_{j}^{-}=-G_{j}^{+},
\end{equation}
where we note $d_{j}^{-}$, $F_{j}^{-}$, $D_{j}^{-}$, $G_{j}^{-}$ corresponding to $-\lambda_j$, and $d_{j}^{+}$, $F_{j}^{+}$, $D_{j}^{+}$, $G_{j}^{+}$ corresponding to $\lambda_j$. Next we will explore the map $\lambda_j$ to $\bar{\lambda}_j$ for those new norming constants. From Proposition 2.1, we have
$$
M'_{n,2}\left( \lambda \right) =\left( \begin{matrix}
	0&		\mp 1\\
	-1&		0\\
\end{matrix} \right) \lambda ^{-2}M'_{n,1}\left( \lambda ^{-1} \right) ,\ \ N'_{n,1}\left( \lambda \right) =\left( \begin{matrix}
	0&		-1\\
	\mp 1&		0\\
\end{matrix} \right) \lambda ^{-2}N'_{n,2}\left( \lambda ^{-1} \right) ,
$$
which can be used to transform Eq.~(2.22b) to another relationship among $M'_{n,1}\left( \lambda _j \right)$, $N_{n,2}\left( \lambda _j \right)$ and $N'_{n,2}\left( \lambda _j \right)$
$$
M'_{n,1}\left( \lambda _j \right) =\mp \left( 2n\bar{b}_j\lambda _{j}^{-2n-1}+\bar{d}_j\lambda _{j}^{2n-2} \right) N_{n,2}\left( \lambda _j \right) \pm \bar{b}_j\lambda _{j}^{-2n}N'_{n,2}\left( \lambda _j \right) .
$$
If we compare above equation with Eq.~(2.22a), we get the second symmetry property for $d_j$ as
\begin{equation}\tag{2.25}
\bar{d}_j=\mp d_j.
\end{equation}
In addition, the relationship between the second and third derivatives of two elements on the main diagonal of the scattering matrix with respect to the spectral parameter can be expressed as
$$
\bar{a}\left( \bar{\lambda}_j \right) =\bar{a}'\left( \bar{\lambda}_j \right) =0,\ \ \bar{a}''\left( \bar{\lambda}_j \right) =\lambda _{j}^{4}a''\left( \lambda _j \right) ,\ \ \bar{a}'''\left( \bar{\lambda}_j \right) =-6\lambda _{j}^{5}a''\left( \lambda _j \right) -\lambda _{j}^{6}a'''\left( \lambda _j \right) ,
$$
this equation, together with Eq.~(2.25), determine the following symmetries
\begin{equation}\tag{2.26}
\bar{F}_j=\pm \bar{\lambda}_{j}^{4}F_j,\ \ \bar{D}_j=-\lambda _{j}^{2}D_j,\ \ \bar{G}_j=-2\lambda _j-\lambda _{j}^{2}G_j.
\end{equation}

Based on the current conclusion, we will construct the solution for Eq.~(1.2) with two-order poles. Considering the absence of reflection coefficient, similar to Section 2.3, the solution of Eq.~(1.2) takes the following form
\begin{equation}\tag{2.27}
u_n=-\frac{a}{2b}+\frac{\sqrt{\sigma \left( a^2+4b \right)}}{2b}\sum_{j=1}^J{\left[ -2\bar{\lambda}_{j}^{-2}P_{-1}\left( \mu _{n+1,21},\bar{\lambda}_j \right) -4\bar{\lambda}_{j}^{-3}P_{-2}\left( \mu _{n+1,21},\bar{\lambda}_j \right) \right]}.
\end{equation}
Similar to the case of one-order poles, we can construct the following system of equations
$$
\left\{ \begin{array}{l}
	-F_1\lambda _{1}^{-2n}\left[ \left( -2n\lambda _{1}^{-1}+D_1-G_1 \right) \left( \frac{1}{\bar{\lambda}_1-\lambda _1}-\frac{1}{\bar{\lambda}_1+\lambda _1} \right) +\frac{1}{\left( \bar{\lambda}_1-\lambda _1 \right) ^2}+\frac{1}{\left( \bar{\lambda}_1+\lambda _1 \right) ^2} \right] \hat{N}_{n,21}\left( \lambda _1 \right)\\
	\ \ \ \ \ \ \ \ \ \ \ \ \ \ \ \ \ \ \ \ \ \ \ \ \ \ \ \ \ \ \ \ \ \ \ \ \ \ \ \ \ \ \ \ \ \ \ \ \ \ \ \ \ +\hat{N}_{n,11}\left( \bar{\lambda}_1 \right) -F_1\lambda _{1}^{-2n}\left( \frac{1}{\bar{\lambda}_1-\lambda _1}-\frac{1}{\bar{\lambda}_1+\lambda _1} \right) \hat{N}_{n,21}'\left( \lambda _1 \right) =1,\\
	-F_1\lambda _{1}^{-2n}\left[ \left( -2n\lambda _{1}^{-1}+D_1-G_1 \right) \left( -\frac{1}{\left( \bar{\lambda}_1-\lambda _1 \right) ^2}+\frac{1}{\left( \bar{\lambda}_1+\lambda _1 \right) ^2} \right) -\frac{2}{\left( \bar{\lambda}_1-\lambda _1 \right) ^3}-\frac{2}{\left( \bar{\lambda}_1+\lambda _1 \right) ^3} \right] \hat{N}_{n,21}\left( \lambda _1 \right)\\
	\ \ \ \ \ \ \ \ \ \ \ \ \ \ \ \ \ \ \ \ \ \ \ \ \ \ \ \ \ \ \ \ \ \ \ \ \ \ \ \ \ \ \ -F_1\lambda _{1}^{-2n}\left( -\frac{1}{\left( \bar{\lambda}_1-\lambda _1 \right) ^2}+\frac{1}{\left( \bar{\lambda}_1+\lambda _1 \right) ^2} \right) \hat{N}_{n,21}'\left( \lambda _1 \right) +\hat{N}_{n,11}'\left( \bar{\lambda}_1 \right) =0,\\
	 -\bar{F}_1\bar{\lambda}_{1}^{2n}\left[ \left( 2n\bar{\lambda}_{1}^{-1}+\bar{D}_1-\bar{G}_1 \right) \left( \frac{1}{\lambda _1-\bar{\lambda}_1}+\frac{1}{\lambda _1+\bar{\lambda}_1} \right) +\frac{1}{\left( \lambda _1-\bar{\lambda}_1 \right) ^2}-\frac{1}{\left( \lambda _1+\bar{\lambda}_1 \right) ^2} \right] \hat{N}_{n,11}\left( \bar{\lambda}_1 \right)\\
	\ \ \ \ \ \ \ \ \ \ \ \ \ \ \ \ \ \ \ \ \ \ \ \  \ \ \ \ \ \ \ \ \ \ \ \ \ \ \ \ \ \ \ \ \ \ \ \ \ \ \ \ \ \ \ +\hat{N}_{n,21}\left( \lambda _1 \right)-\bar{F}_1\bar{\lambda}_{1}^{2n}\left( \frac{1}{\lambda _1-\bar{\lambda}_1}+\frac{1}{\lambda _1+\bar{\lambda}_1} \right) \hat{N}_{n,11}'\left( \bar{\lambda}_1 \right) =0,\\
	-\bar{F}_1\bar{\lambda}_{1}^{2n}\left[ \left( 2n\bar{\lambda}_{1}^{-1}+\bar{D}_1-\bar{G}_1 \right) \left( -\frac{1}{\left( \lambda _1-\bar{\lambda}_1 \right) ^2}-\frac{1}{\left( \lambda _1+\bar{\lambda}_1 \right) ^2} \right) -\frac{2}{\left( \lambda _1-\bar{\lambda}_1 \right) ^3}+\frac{2}{\left( \lambda _1+\bar{\lambda}_1 \right) ^3} \right] \hat{N}_{n,11}\left( \bar{\lambda}_1 \right)\\
	\ \ \ \ \ \ \ \ \ \ \ \ \ \ \ \ \ \ \ \ \ \ \ \ \ \ \ \ \ \ \ \ \ \ \ \ \ \ \ \ \ \ \ \  \ +\hat{N}_{n,21}'\left( \lambda _1 \right) -\bar{F}_1\bar{\lambda}_{1}^{2n}\left( -\frac{1}{\left( \lambda _1-\bar{\lambda}_1 \right) ^2}-\frac{1}{\left( \lambda _1+\bar{\lambda}_1 \right) ^2} \right) \hat{N}_{n,11}'\left( \bar{\lambda}_1 \right) =0,\\
\end{array} \right.
$$
to obtain the unknown variables in Eq.~(2.27) if we only take into account a set of discrete eigenvalues. The final step in completing the solution is to consider the time evolution of the norming constant. Considering Eqs.~(2.19) and (2.22), taking derivative of $\hat{M}_{n}\left( \lambda ,t \right)$ and evaluating it on $\lambda_j$, we have
\begin{align}\notag
\hat{M}_{n,1}^{\left( 1,0 \right)}\left( \lambda _j,t \right) =&\lambda _{j}^{-2n}e^{\left( \omega _1\left( \lambda _j \right) -\omega _2\left( \lambda _j \right) \right) t}\\\notag
&\left( \left[ \left( \omega _1'\left( \lambda _j \right) -\omega _2'\left( \lambda _j \right) \right) tb_j\left( t \right) -2n\lambda _{j}^{-1}b_j\left( t \right) +d_j\left( t \right) \right] \hat{N}_{n,2}\left( \lambda _j,t \right) +b_j\left( t \right) \hat{N}_{n,2}^{\left( 1,0 \right)}\left( \lambda _j,t \right) \right) .\notag
\end{align}
Recalling that $\hat{M}_{n,1}\left( \lambda ,t \right)$ simultaneously satisfies Lax pair (2.1), one can obtain the following equations
\begin{align}\notag
\hat{M}_{n,1}^{\left( 1,1 \right)}\left( \lambda _j,t \right) =&\lambda _{j}^{-2n}e^{\left( \omega _1\left( \lambda _j \right) -\omega _2\left( \lambda _j \right) \right) t}\\ \notag
&\left( \left[ d_j\left( t \right) \left( \omega _1\left( \lambda _j \right) -\omega _2\left( \lambda _j \right) \right) +\left( \omega _1'\left( \lambda _j \right) -\omega _2'\left( \lambda _j \right) \right) b_j\left( t \right) +d_j'\left( t \right) \right] \hat{N}_{n,2}\left( \lambda _j,t \right) \right.\\\notag
&\left.+\left[ \left( \omega _1'\left( \lambda _j \right) -\omega _2'\left( \lambda _j \right) \right) tb_j\left( t \right) -2n\lambda _{j}^{-1}b_j\left( t \right) +d_j\left( t \right) \right] \hat{N}_{n,2}^{\left( 0,1 \right)}\left( \lambda _j,t \right) +b_j\left( t \right) \hat{N}_{n,2}^{\left( 1,1 \right)}\left( \lambda _j,t \right) \right) ,\\ \notag
\hat{M}_{n,1}^{\left( 1,1 \right)}\left( \lambda _j,t \right) =&\lambda _{j}^{-2n}e^{\left( \omega _1\left( \lambda _j \right) -\omega _2\left( \lambda _j \right) \right) t}\\ \notag
&\left( \left[ \left( \omega _1'\left( \lambda _j \right) -\omega _2'\left( \lambda _j \right) \right) tb_j\left( t \right) -2n\lambda _{j}^{-1}b_j\left( t \right) +d_j\left( t \right) \right] \hat{N}_{n,2}^{\left( 0,1 \right)}\left( \lambda _j,t \right) +b_j\left( t \right) \hat{N}_{n,2}^{\left( 1,1 \right)}\left( \lambda _j,t \right) \right) ,\notag
\end{align}
resulting
\begin{equation}\tag{2.28}
d_j\left( t \right) =e^{\left( \omega _2\left( \lambda _j \right) -\omega _1\left( \lambda _j \right) \right) t}\left[ d_j\left( 0 \right) +b_j\left( 0 \right) t\left( \omega _2'\left( \lambda _j \right) -\omega _1'\left( \lambda _j \right) \right) \right] .
\end{equation}
Right now, we have successfully constructed the two-order pole solution (see Fig.~7) of Eq.~(1.2) through IST.
\begin{center}
\includegraphics[scale=0.43]{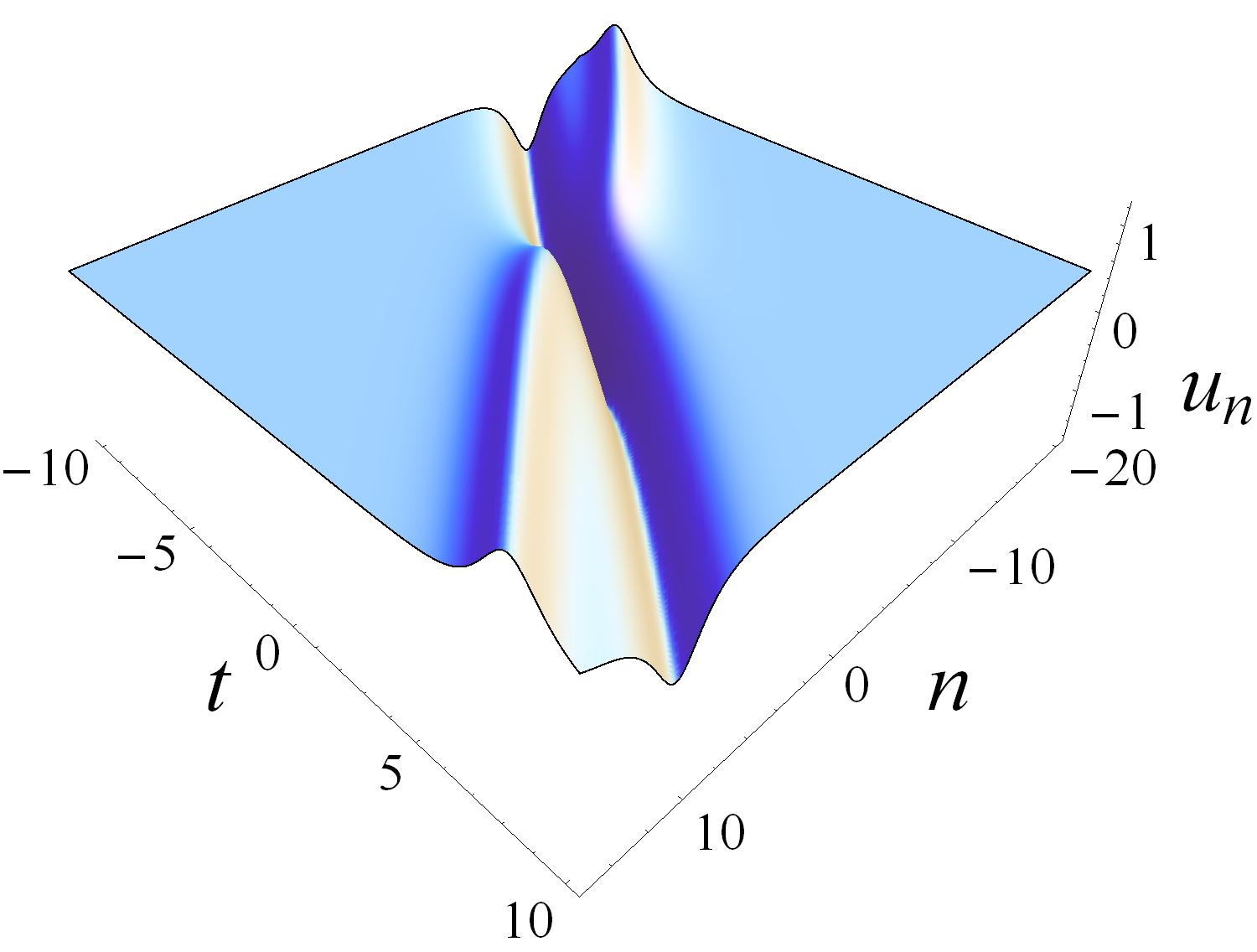}\hfill
\includegraphics[scale=0.43]{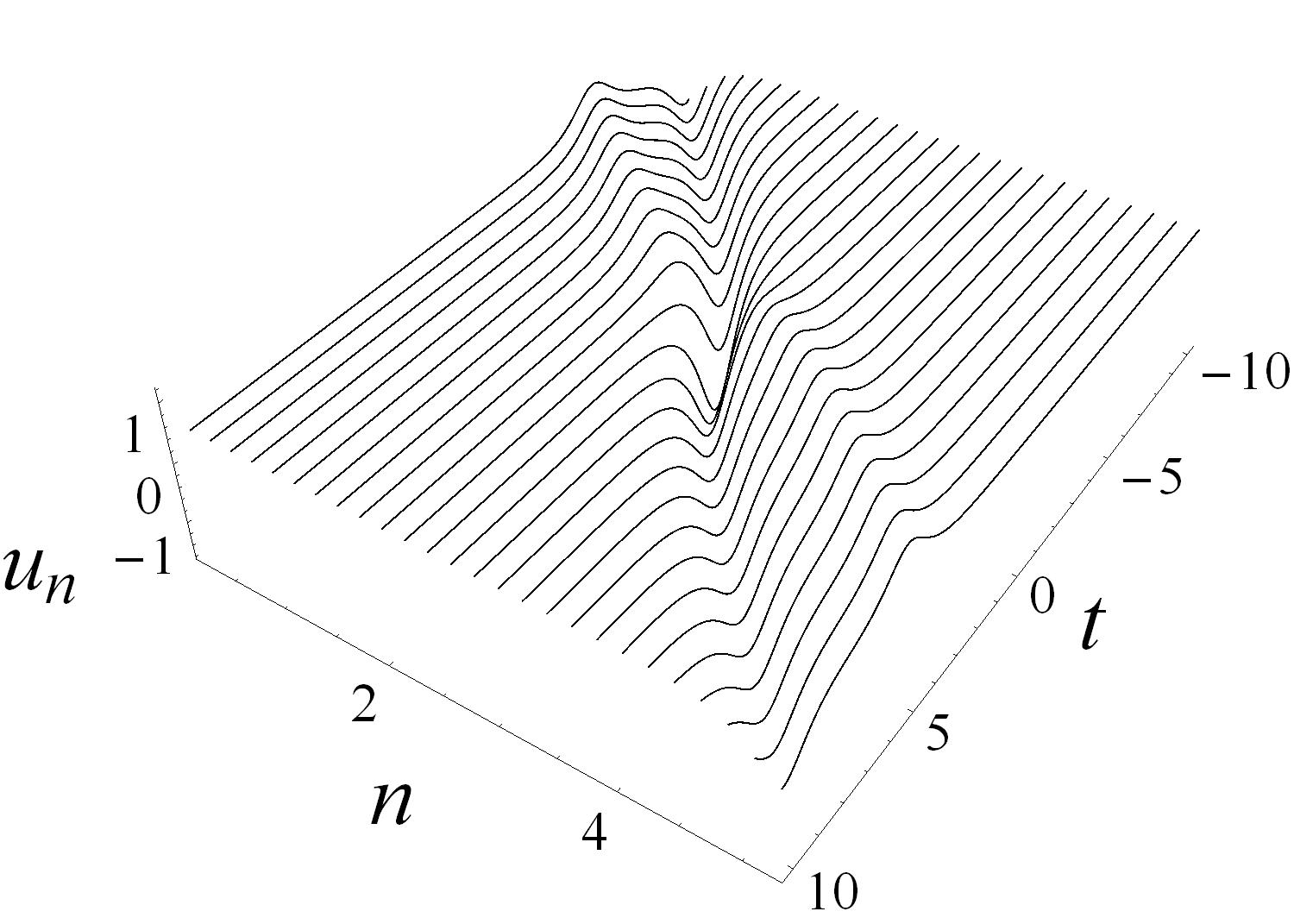}\hfill
\includegraphics[scale=0.35]{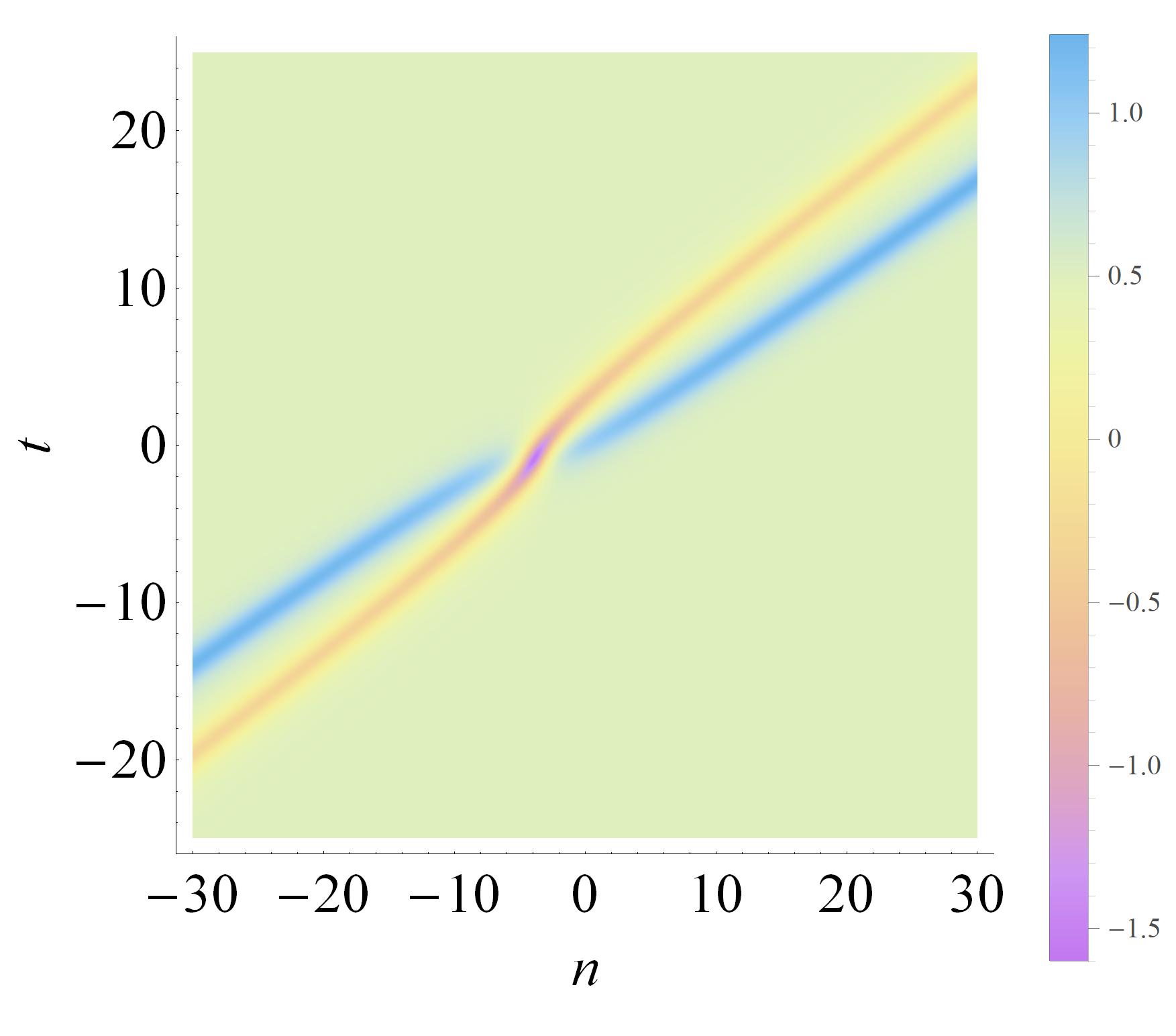}\hfill\\
\vspace{-0.2cm}{\footnotesize\hspace{0cm}(a)\hspace{5.9cm}(b)\hspace{5.9cm}(c)}
\flushleft{\footnotesize
\textbf{Fig.~$7$.} Solutions corresponding to Eq.~(2.27), the two-order solution with $a=1$, $b=-1$, $\lambda_1=1.5$, $b_1(0)=-1$, $d_1(0)=-1$.}
\end{center}
Easy to see, it does not produce large amplitude at the moment of collision as the two-soliton collision in Section 2.3.

\vspace{5mm}\noindent\textbf{2.5 Trace formula }

For the case in Section 2.3, we introduce
$$
\alpha \left( \lambda \right) =\prod_{j=1}^J{\frac{\lambda ^2-\lambda _{j}^{-2}}{\lambda ^2-\lambda _{j}^{2}}a\left( \lambda \right)},\ \ \bar{\alpha}\left( \lambda \right) =\prod_{j=1}^J{\frac{\lambda ^2-\bar{\lambda}_{j}^{-2}}{\lambda ^2-\bar{\lambda}_{j}^{2}}a\left( \lambda \right)}
$$
to remove the effect of zeros. Based on the analyticity of scattering coefficients, we have
$$
\text{log}\left[ a\left( \lambda \right) \right] =\sum_{j=1}^J{\text{log}\left[ \frac{\lambda ^2-\lambda _{j}^{2}}{\lambda ^2-\lambda _{j}^{-2}} \right] -\frac{1}{2\pi i}\oint_{\left| \omega \right|=1}{\frac{\omega \text{log}\left[ \alpha \left( \omega \right) \bar{\alpha}\left( \omega \right) \right]}{\omega ^2-\lambda ^2}d\omega}}\ \ \ \ \ \ \ \ \ \ \text{as}\left| \lambda \right|>1,
$$
$$
\text{log}\left[ \bar{a}\left( \lambda \right) \right] =\sum_{j=1}^J{\text{log}\left[ \frac{\lambda ^2-\bar{\lambda}_{j}^{2}}{\lambda ^2-\bar{\lambda}_{j}^{-2}} \right] +\frac{1}{2\pi i}\oint_{\left| \omega \right|=1}{\frac{\omega \text{log}\left[ \alpha \left( \omega \right) \bar{\alpha}\left( \omega \right) \right]}{\omega ^2-\lambda ^2}d\omega}}\,\,\,\,\,\,\,\,\,\,\,\,\,\,\,\,\,\,\,\,\text{as}\left| \lambda \right|<1.
$$
If the reflection vanishes and there exists only a set of discrete eigenvalues, the scattering coefficients can be rewritten as
$$
a\left( \lambda \right) =\frac{\lambda ^2-\lambda _{1}^{2}}{\lambda ^2-\lambda _{1}^{-2}},\ \ \bar{\alpha}\left( \lambda \right) =\frac{\lambda ^2-\bar{\lambda}_{1}^{2}}{\lambda ^2-\bar{\lambda}_{1}^{-2}},
$$
which implies
$$
a'\left( \lambda _1 \right) =\frac{2\lambda _{1}^{3}}{\lambda _{1}^{4}-1}.
$$
Also, for the case of two-order poles, we get the following results
$$
a''\left( \lambda _1 \right) =\frac{8\lambda _{1}^{2}}{\left( \lambda _{1}^{2}-\bar{\lambda}_{1}^{2} \right) ^2},\ a'''\left( \lambda _1 \right) =-\frac{24\lambda _1\left( 3\lambda _{1}^{2}+\bar{\lambda}_{1}^{2} \right)}{\left( \lambda _{1}^{2}-\bar{\lambda}_{1}^{2} \right) ^3},
$$
which can be used in Eq.~(2.27).

\vspace{5mm} \noindent\textbf{3  The specific step-like boundary condition, \bm{$u_n$} tends to \bm{$\frac{c_{\pm}\sqrt{ a^2+4b }-a}{2b}$} as \bm{$n\rightarrow \pm \infty $}}
\hspace*{\parindent}
\renewcommand{\theequation}{3.\arabic{equation}}\setcounter{equation}{0}

In this section, we focus on a special step-like problem with the case of $\sigma=1$ and the following boundary condition
\begin{equation}\tag{3.1}
u_n\rightarrow u_{\pm}=\frac{c_{\pm}\sqrt{a^2+4b}-a}{2b},\ \ \text{as}\ n\rightarrow \pm \infty
\end{equation}
in which $\left| c_{\pm} \right|=c_0=const$. Then the scattering problem satisfies
\begin{equation}\tag{3.2}
\psi _{n+1}\rightarrow \left( \begin{matrix}
	\lambda&		c_{\pm}\\
	c_{\pm}&		\lambda ^{-1}\\
\end{matrix} \right) \psi _n
\end{equation}
which implies that
\begin{equation}\tag{3.3}
c_{\pm}\psi _{n}^{\left( 2 \right)}=\psi _{n+1}^{\left( 1 \right)}-\lambda \psi _{n}^{\left( 1 \right)},\ \ \
c_{\pm}\psi _{n}^{\left( 1 \right)}=\psi _{n+1}^{\left( 2 \right)}-\lambda^{-1} \psi _{n}^{\left( 2 \right)}.
\end{equation}

\vspace{5mm}\noindent\textbf{3.1 Jost eigenfunctions and scattering coefficients }

From Eq.~(3.2) we can define the Jost eigenfunctions according to their asymptotic behaviors at large $n$ as follows
\begin{equation}\tag{3.4a}
\varGamma _n\left( \lambda \right) \rightarrow r^n\left( \begin{matrix}
	k^{-n}\left( kr-\lambda \right)&k^nc_+\\
-k^{-n}c_+&		k^n\left( kr-\lambda \right)\\
\end{matrix} \right) \ \ \ \ \ n\rightarrow +\infty ,
\end{equation}
\begin{equation}\tag{3.4b}
\varLambda _n\left( \lambda \right) \rightarrow r^n\left( \begin{matrix}
	k^{-n}\left( kr-\lambda \right)&k^nc_-\\
-k^{-n}c_-&		k^n\left( kr-\lambda \right)\\
\end{matrix} \right) \,\,\,\,\,\,\,\,\,\,n\rightarrow -\infty ,
\end{equation}
where
$$
r=\sqrt{1-c_{0}^{2}},\ \ r\left( k+k^{-1} \right) =\lambda +\lambda ^{-1}.
$$
Note that, we choose $c_0<1$ here such that $r\in \mathbb{R}$. Easy to see
$$
k\left( \lambda \right) =\frac{\lambda +\lambda ^{-1}}{2r}\pm \sqrt{\left( \frac{\lambda +\lambda ^{-1}}{2r} \right) ^2-1},\ \ k^{-1}\left( \lambda \right) =\frac{\lambda +\lambda ^{-1}}{2r}\mp \sqrt{\left( \frac{\lambda +\lambda ^{-1}}{2r} \right) ^2-1},
$$
implying that there are four branch points $\pm r\pm ic_0$ on the $\lambda$-plane located on the unite circle. The existence of these branch points leads to multivaluability, which can be removed by introducing new variable $\zeta=k/\lambda$. Thus, any quantities which can be expressed as even functions of $\lambda$ and $k$ are all meromorphic functions of $\zeta$. For any two solutions $\nu _n$, $\omega _n$ for the scattering problem (1.3), their Wronskian satisfies the iterative relationship $
Wr\left( \nu _{n+1},\omega _{n+1} \right) =\left( 1-\frac{\left( 2bu_n+a \right) ^2}{a^2+4b} \right) Wr\left( \nu _n,\omega _n \right)$. In addition, the Wronskian of two Jost eigenfunction tends to $
r^{2n}\left( \left( kr-\lambda \right) ^2+c_{0}^{2} \right)$ at the boundary, respectively. These two conditions determine
\begin{equation}\tag{3.5}
det \left( \varGamma _n\left( \lambda \right)  \right) =\left[ \left( kr-\lambda \right) ^2+c_{0}^{2}  \right] r^{2n}\prod_{j=-\infty}^{n-1}{\frac{1-\frac{\left( 2bu_j+a \right) ^2}{a^2+4b}}{1-c_{0}^{2}}},$$
$$
det\left( \varLambda _n\left( \lambda \right) \right) =\left[ \left( kr-\lambda \right) ^2+c_{0}^{2}  \right] r^{2n}\prod_{j=n}^{+\infty}{\frac{1-c_{0}^{2}}{1-\frac{\left( 2bu_j+a \right) ^2}{a^2+4b}}}.
\end{equation}
Note that
$$
\left( kr-\lambda \right) ^2+c_{0}^{2}=-\left( kr-\lambda \right) \left( \frac{1}{kr}-\frac{1}{\lambda} \right) +c_{0}^{2}=-2r^2+r\left( \frac{k}{\lambda}+\frac{\lambda}{k} \right) =r\left( \zeta +\zeta ^{-1}-2r \right) \ne 0,
$$
this indicates that $\varGamma_n(\lambda)$ and $\varLambda_n(\lambda)$ are two fundamental solutions of scattering problem i.e., there exists a scattering matrix $T\left( \lambda \right) =\left( \begin{matrix}
	\varsigma \left( \lambda \right)&		\bar{\tau}\left( \lambda \right)\\
	\tau \left( \lambda \right)&		\bar{\varsigma}\left( \lambda \right)\\
\end{matrix} \right)$ that is independent of $n$ such that
$
\varLambda _n\left( \lambda \right) =\varGamma _n\left( \lambda \right) T\left( \lambda \right),
$
then,
$$
\varsigma \left( \lambda \right) =\frac{Wr\left( \varLambda _{n,2}\left( \lambda \right) ,\varGamma _{n,1}\left( \lambda \right) \right)}{Wr\left( \varGamma _{n,2}\left( \lambda \right) ,\varGamma _{n,1}\left( \lambda \right) \right)},\ \bar{\varsigma}\left( \lambda \right) =-\frac{Wr\left( \varLambda _{n,1}\left( \lambda \right) ,\varGamma _{n,2}\left( \lambda \right) \right)}{Wr\left( \varGamma _{n,2}\left( \lambda \right) ,\varGamma _{n,1}\left( \lambda \right) \right)},
$$
$$
\tau \left( \lambda \right) =-\frac{Wr\left( \varLambda _{n,2}\left( \lambda \right) ,\varGamma _{n,2}\left( \lambda \right) \right)}{Wr\left( \varGamma _{n,2}\left( \lambda \right) ,\varGamma _{n,1}\left( \lambda \right) \right)},\ \ \bar{\tau}\left( \lambda \right) =\frac{Wr\left( \varLambda _{n,1}\left( \lambda \right) ,\varGamma _{n,1}\left( \lambda \right) \right)}{Wr\left( \varGamma _{n,2}\left( \lambda \right) ,\varGamma _{n,1}\left( \lambda \right) \right)}.
$$

In order to construct modified eigenfunctions which are independent of $k$ when $n\rightarrow \pm \infty $, we introduce two auxiliary matrices
\begin{equation}\notag
\varDelta =\left( \begin{matrix}
	k&		0\\
	0&		\frac{1}{k}\\
\end{matrix} \right) ,\ \ \ \ \ \ \varOmega =\left( \begin{matrix}
	1&		0\\
	0&		k\\
\end{matrix} \right) .
\end{equation}
Then we can define modified eigenfunctions as
\begin{equation}\tag{3.6a}
X_n\left( \zeta \right) =r^{-n}\varOmega \left( k \right) \varLambda _n\left( \lambda \right) \varOmega \left( k \right) ^{-1}\varDelta \left( k \right) ^{-n}\rightarrow \left( \begin{matrix}
	r-\zeta ^{-1}&		c_-\\
	-c_-&		\zeta -r\\
\end{matrix} \right) ,\ \ \ n\rightarrow -\infty ,
\end{equation}
\begin{equation}\tag{3.6b}
Y_n\left( \zeta \right) =r^{-n}\varOmega \left( k \right) \varGamma _n\left( \lambda \right) \varOmega \left( k \right) ^{-1}\varDelta \left( k \right) ^{-n}\rightarrow \left( \begin{matrix}
	r-\zeta ^{-1}&		c_+\\
	-c_+&		\zeta -r\\
\end{matrix} \right) ,\ \ \ n\rightarrow +\infty ,
\end{equation}
and they satisfy the following iterative relationship
$$
rX_{n+1,1}\left( \zeta \right) =B_nX_{n,1}\left( \zeta \right) ,\ rX_{n+1,2}\left( \zeta \right) =A_nX_{n,2}\left( \zeta \right) ,
$$
$$
rY_{n+1,1}\left( \zeta \right) =B_nY_{n,1}\left( \zeta \right) ,\,\,rY_{n+1,2}\left( \zeta \right) =A_nY_{n,2}\left( \zeta \right) ,
$$
where
$$
A_n=\left( \begin{matrix}
	\frac{\lambda}{k}&		\frac{2bu_n+a}{k^2\sqrt{a^2+4b}}\\
	\frac{2bu_n+a}{\sqrt{a^2+4b}}&		\frac{1}{\lambda k}\\
\end{matrix} \right) ,\ \ B_n=\left( \begin{matrix}
	\lambda k&		\frac{2bu_n+a}{\sqrt{a^2+4b}}\\
	\frac{k^2\left( 2bu_n+a \right)}{\sqrt{a^2+4b}}&		\frac{k}{\lambda}\\
\end{matrix} \right) .
$$
\\
\textbf{Proposition 3.1} {\it It can be affirmed that the modified eigenfunctions defined in Eqs.~(3.6) that satisfy specific boundary conditions have the following properties:

(1) Analyticity. Using the Neumann series, we have come to the conclusion that $X_{n,2}\left( \zeta \right)$, $Y_{n,1}\left( \zeta \right)$ are analytic outside the unit circle $\left| \zeta \right|= 1$ except for infinities and the branch points, while $X_{n,1}\left( \zeta \right)$, $Y_{n,2}\left( \zeta \right)$ are analytic in the region of $\left| \zeta \right|< 1$ excluding the branch points and origin, and the four eigenfunctions are all continuous on the unit circle. Further more, $\varsigma(\zeta )$ is analytic outside the unit circle while $\bar{\varsigma}(\zeta)$ is analytic inside the unit circle, and the other scattering coefficients only can be defined on the unit circle.

 (2) Asymptotic behaviors. From the relationship among $\zeta$, $\lambda$ and $k$, we find that when $\zeta \rightarrow r$, one have $\lambda \rightarrow 0$ and $k \rightarrow 0$, when $\zeta \rightarrow \frac{1}{r}$, one have $\lambda \rightarrow \infty$ and $k \rightarrow \infty$. This indicates that $\zeta =r$ and $\zeta = \frac{1}{r}$ respectively play the same role as $\zeta =0$ and $\zeta =\infty$. At these four singular points, the modified eigenfunctions exhibit the following asymptotic behaviors by the WKB expansion
 \begin{equation}\tag{3.7a}
X_{n,1}\left( \zeta \right) \rightarrow -c_{-}\left( \begin{array}{c}
	\frac{2bu_{n-1}+a}{r\sqrt{a^2+4b}}\\
	1\\
\end{array} \right),\ \ \ Y_{n,2}\left( \zeta \right) \rightarrow \frac{c_+}{\varDelta _n}\left( \begin{array}{c}
	1\\
	\frac{2bu_n+a}{c_{0}^{2}\sqrt{a^2+4b}}\left( \zeta -r \right)\\
\end{array} \right) \ \ \zeta \rightarrow r,
\end{equation}
\begin{equation}\tag{3.7b}
X_{n,2}\left( \zeta \right) \rightarrow \left( \begin{array}{c}
	\frac{2bu_{n-1}+a}{\sqrt{a^2+4b}}\\
	\zeta\\
\end{array} \right) ,\ \ \ Y_{n,1}\left( \zeta \right) \rightarrow \frac{1}{\varDelta _n}\left( \begin{array}{c}
	r\\
	-\frac{2bu_n+a}{\sqrt{a^2+4b}}\\
\end{array} \right)\ \ \ \ \ \ \ \ \ \ \ \ \ \zeta \rightarrow \infty,
\end{equation}
\begin{equation}\tag{3.7c}
Y_{n,2}\left( \zeta \right) \rightarrow \frac{1}{\varDelta _n}\left( \begin{array}{c}
	\frac{2bu_n+a}{\sqrt{a^2+4b}}\\
	-r\\
\end{array} \right) ,\ \ \ X_{n,1}\left( \zeta \right) \rightarrow \left( \begin{array}{c}
	-\frac{1}{\zeta}\\
	-\frac{2bu_{n-1}+a}{\sqrt{a^2+4b}}\\
\end{array} \right)\ \ \ \ \ \ \ \ \ \ \ \ \ \zeta \rightarrow 0,
\end{equation}
\begin{equation}\tag{3.7d}
X_{n,2}\left( \zeta \right) \rightarrow c_-\left( \begin{array}{c}
	1\\
	\frac{2bu_{n-1}+a}{r\sqrt{a^2+4b}}\\
\end{array} \right) ,\ \ \ Y_{n,1}\left( \zeta \right) \rightarrow -\frac{c_{+}}{\varDelta _n}\left( \begin{array}{c}
	\frac{r^2 (2bu_n+a)}{c_{0}^{2}\sqrt{a^2+4b}}\left( \zeta -\frac{1}{r} \right)\notag\\
	1\\
\end{array} \right) \ \ \ \ \ \zeta \rightarrow \frac{1}{r},
\end{equation}
for $
\varDelta _n=\prod_{j=n}^{+\infty}{\dfrac{\left( 1-\frac{\left( 2bu_j+a \right) ^2}{a^2+4b} \right)}{\left( 1-c_{0}^{2} \right)}}
$. Based on the relations between eigenfunctions and scattering coefficients, it can be further obtained that
\begin{equation}\tag{3.8a}
\varsigma \left( \zeta \right) \rightarrow 1\quad\quad\zeta \rightarrow \infty ,\quad\quad\quad
\bar{\varsigma }\left( \zeta \right) \rightarrow 1\quad\quad\zeta \rightarrow 0,
\end{equation}
\begin{equation}\tag{3.8b}
\bar{\varsigma }\left( \zeta \right) \rightarrow \frac{c_{-}c_+}{1-r^2}\quad\quad\zeta \rightarrow r,\quad\quad\quad
\varsigma \left( \zeta \right) \rightarrow \frac{c_-c_{+}}{1-r^2}\quad\quad\zeta \rightarrow \frac{1}{r}.
\end{equation}

(3) Symmetry 1: $\left( \lambda,k \right) \rightarrow \left( \lambda^{-1},k^{-1} \right)$.

For $
\sigma _1=\left( \begin{matrix}
	0&		1\\
	1&		0\\
\end{matrix} \right)$, the Jost eigenfunctions satisfy
\begin{equation}\tag{3.9}
\varLambda _n\left( \lambda ,k \right) =-\sigma _1\varLambda _n\left( \lambda ^{-1},k^{-1} \right) \sigma _1,\ \varGamma _n\left( \lambda ,k \right) =-\sigma _1\varGamma _n\left( \lambda ^{-1},k^{-1} \right) \sigma _1,
\end{equation}
and the scattering matrix satisfies
\begin{equation}\tag{3.10}
T\left( \lambda ,k \right) =\sigma _1T_n\left( \lambda ^{-1},k^{-1} \right) \sigma _1,
\end{equation}
which indicates that for a zero $\zeta _j$ of $\varsigma(\zeta)$, there exists $\bar{\zeta}_j=\zeta _{j}^{-1}$ such that $\bar{\varsigma}(\bar{\zeta}_j)=0$, vice versa, i.e., the discrete eigenvalues inside and outside the unit circle appear in pairs.

(4) Symmetry 2: $\left( \lambda,k \right) \rightarrow \left( \lambda,k^{-1} \right)$.

Due to the fact that the Jost eigenfunctions all satisfy the scattering problem which is independent of $k$, by considering the asymptotic behaviors of them, we obtain
\begin{equation}\tag{3.11}
\varLambda _{n,2}\left( \lambda ,k^{-1} \right) =\frac{\lambda -rk^{-1}}{c_-}\varLambda _{n,1}\left( \lambda ,k \right) ,\ \ \varGamma _{n,1}\left( \lambda ,k^{-1} \right) =\frac{rk^{-1}-\lambda}{c_+}\varGamma _{n,2}\left( \lambda ,k^{-1} \right) ,
\end{equation}
\begin{equation}\tag{3.12a}
\tau \left( \lambda ,k \right) =-\frac{c_-}{c_+}\tau \left( \lambda ,k^{-1} \right) ,\ \ \varsigma \left( \lambda ,k \right) =\frac{c_-}{c_+}\varsigma \left( \lambda ,k^{-1} \right) ,
\end{equation}
\begin{equation}\tag{3.12b}
\bar{\tau}\left( \lambda ,k \right) =-\frac{c_-}{c_+}\bar{\tau}\left( \lambda ,k^{-1} \right) ,\,\,\,\,\bar{\varsigma}\left( \lambda ,k \right) =\frac{c_-}{c_+}\bar{\varsigma}\left( \lambda ,k^{-1} \right) .
\end{equation}
Combining the two symmetries, one yields
\begin{equation}\tag{3.13}
\varLambda _n\left( \lambda ,k^{-1} \right) =\frac{\lambda -rk^{-1}}{c_-}\sigma _1\varLambda _n\left( \lambda ^{-1},k^{-1} \right) \sigma _3,\ \
\varGamma _n\left( \lambda ,k^{-1} \right) =\frac{\lambda -rk^{-1}}{c_+}\sigma _1\varGamma _n\left( \lambda ^{-1},k^{-1} \right) \sigma _3,
\end{equation}
with $
\sigma _3=\left( \begin{matrix}
	1&		0\\
	0&		-1\\
\end{matrix} \right).
$
}

\vspace{5mm}\noindent\textbf{3.2 Inverse problem }

From Eqs.~(3.6) and definition of $T(\lambda)$, we obtain the jump relationship between two meromorphic matrix functions on $|\zeta|=1$
\begin{equation}\tag{3.14}
\left( \begin{matrix}
	Y_{n,1}\left( \zeta \right)&		\frac{X_{n,2}\left( \zeta \right)}{\varsigma \left( \zeta \right)}\\
\end{matrix} \right) =\left( \begin{matrix}
	\frac{X_{n,1}\left( \zeta \right)}{\bar{\varsigma}\left( \zeta \right)}&		Y_{n,2}\left( \zeta \right)\\
\end{matrix} \right) \left( \begin{matrix}
	1&		k^{-2n+1}\frac{\tau \left( \zeta \right)}{\varsigma \left( \zeta \right)}\\
	-k^{2n-1}\frac{\bar{\tau}\left( \zeta \right)}{\bar{\varsigma}\left( \zeta \right)}&		1-\frac{\tau \left( \zeta \right)}{\varsigma \left( \zeta \right)}\frac{\bar{\tau}\left( \zeta \right)}{\bar{\varsigma}\left( \zeta \right)}\\
\end{matrix} \right).
\end{equation}
The reason why two functions composed of modified eigenfunctions in the above equation are meromorphic is due to the existence of zeros of $\varsigma (\zeta)$ and $\bar{\varsigma} (\zeta)$, which lead to solitons and can be proven to be simple.

Similar to Section 2.5, we can derive the trace formula for $J$ simple zeros $\zeta_j$ of $\bar{\varsigma} (\zeta)$
$$
\bar{\varsigma}\left( \zeta \right) =\prod_{j=1}^J{\bar{\zeta}_j\frac{\zeta -\bar{\zeta}_j}{\zeta \bar{\zeta}_j-1}}\exp \left[ \frac{1}{2\pi i}\oint_{\left| \omega \right|=1}{\frac{\text{log}\left| \varsigma \left( \omega \right) \right|^2}{\omega -\zeta}d\omega} \right] ,
$$
and the theta condition
$$
\frac{c_+c_-}{1-r^2}=\prod_{j=1}^J{\bar{\zeta}_j\frac{r-\bar{\zeta}_j}{r\bar{\zeta}_j-1}}\exp \left[ \frac{1}{2\pi i}\oint_{\left| \omega \right|=1}{\frac{\text{log}\left| \varsigma \left( \omega \right) \right|^2}{\omega -\zeta}d\omega} \right] ,
$$
considering Eqs.~(3.8b). It has been shown that if $c_+=c_-$, in the case of no reflection and only one set of discrete eigenvalues, the zero of $\bar{\varsigma} (\zeta)$ must satisfies $\left|\bar{\zeta}_j\right|=1$, i.e., along the continuous spectrum. As a consequence, the corresponding eigenfunctions don't decay as $x\rightarrow\pm\infty$ and no bound state is produced.

For a zero $\lambda_j$ of $\varsigma (\lambda)$, there is a constant $\tau_j$ such that $\varLambda _{n,2}\left( \lambda _j \right) =\tau _j\varGamma _{n,1}\left( \lambda _j \right)$, while $
\varLambda _{n,1}\left( \bar{\lambda}_j \right) =\bar{\tau}_j\varGamma _{n,2}\left( \bar{\lambda}_j \right)$ for $\bar{\varsigma} (\bar{\lambda}_j)=0$. Naturally, we must consider the residues of these meromorphic functions at their poles i.e., the zeros of $\varsigma(\zeta)$ and $\bar{\varsigma}(\zeta)$ respectively as follows
\begin{equation}\tag{3.15}
\text{Res}\left( \frac{X_{n,2}\left( \zeta \right)}{\varsigma \left( \zeta \right)};\zeta =\zeta _j \right) =C_jk \left( \zeta _j \right) ^{-2n}Y_{n,1}\left( \zeta _j \right) ,$$
$$
\text{Res}\left( \frac{X_{n,1}\left( \zeta \right)}{\bar{\varsigma}\left( \zeta \right)};\zeta =\bar{\zeta}_j \right) =\bar{C}_jk \left( \bar{\zeta}_j \right) ^{2n}Y_{n,2}\left( \bar{\zeta}_j \right) ,
\end{equation}
with $
C_j=\frac{\tau_j\lambda \left( \zeta _j \right)}{\varsigma'\left( \zeta _j \right)}$, $ \bar{C}_j=\frac{\bar{\tau}_j}{\lambda \left( \zeta _j \right) \bar{\varsigma}'\left( \bar{\zeta}_j \right)}$. From Eq.~(3.10), we get $
\varsigma '\left( \zeta _j \right) =-\bar{\zeta}_{j}^{2}\bar{\varsigma}'\left( \bar{\zeta}_j \right)
$, further more, if the potentials $\frac{2bu_n-a}{\sqrt{a^2+4b}}-c_\pm$ decay rapidly enough as $n\rightarrow\pm\infty$, such that $k\tau \left( \zeta \right) $, $\frac{\bar{\tau}\left( \zeta \right)}{k}$ can be extended off the unit circle in correspondence of the discrete eigenvalues, then we can obtain the relation between $C_j$ and $\bar{C}_j$ as
\begin{equation}\tag{3.16}
\bar{C}_j=-\bar{\zeta}_j^2C_j,
\end{equation}
noticing that $k \left( \lambda \right) =k^{-1} \left(\lambda^{-1} \right)$.

Next, we divide the first element of Eqs.~(3.13) by $\zeta -r$
\begin{equation}\tag{3.17}
\frac{X_{n,2}\left( \zeta \right)}{\left( \zeta -r \right) \varsigma\left( \zeta \right)}=\frac{Y_{n,2}\left( \zeta \right)}{\zeta -r}+\frac{k \left( \zeta \right) ^{-2n+1}}{\zeta -r}Y_{n,1}\left( \zeta \right)  \frac{\tau(\zeta)}{\varsigma(\zeta)}.
\end{equation}
Since $r<1$, the LHS is meromorphic outside the unit circle and tends to $\left( \begin{array}{c}
	0\\
	1\\
\end{array} \right)
$ from Eqs.~(3.7b) and (3.8a). To avoid the influence of singularities, we remove the asymptotic behavior at infinity and the residue effect at zeros of $\varsigma(\zeta)$ on both sides of Eq.~(3.17)
\begin{align}
\frac{X_{n,2}\left( \zeta \right)}{\left( \zeta -r \right) \varsigma\left( \zeta \right)}-&\left( \begin{array}{c}
	0\\
	1\\
\end{array} \right) -\sum_{j=1}^J{\frac{\text{Res}\left( \frac{X_{n,2}}{\varsigma};\zeta _j \right)}{\left( \zeta _j-r \right) \left( \zeta -\zeta _j \right)}}\notag\\
&=\frac{Y_{n,2}\left( \zeta \right)}{\zeta -r}-\left( \begin{array}{c}
	0\\
	1\\
\end{array} \right) -\sum_{j=1}^J{\frac{\text{Res}\left( \frac{X_{n,2}}{\varsigma};\zeta _j \right)}{\left( \zeta _j-r \right) \left( \zeta -\zeta _j \right)}}+\frac{k \left( \zeta \right) ^{-2n+1}}{\zeta -r}Y_{n,1}\left( \zeta \right) \frac{\tau(\zeta)}{\varsigma(\zeta)}  .\tag{3.18}
\end{align}
In above equation, the LHS is analytic outside the unit circle and decays as $\zeta \rightarrow \infty$, then we get
\begin{align}
\frac{1}{2\pi i}\int_{\left| \omega \right|=1}{\left[ \frac{Y_{n,2}\left( \omega \right)}{\omega -r}-\left( \begin{array}{c}
	0\\
	1\\
\end{array} \right) -\sum_{j=1}^J{\frac{\text{Res}\left( \frac{X_{n,2}}{\varsigma};\zeta _j \right)}{\left( \zeta _j-r \right) \left( \omega -\zeta _j \right)}}\right.}{\left.+\frac{k \left( \omega \right) ^{-2n+1}}{\omega -r}Y_{n,1}\left( \omega \right) \frac{\tau(\omega)}{\varsigma(\omega)} \right]}\frac{d\omega}{\omega -\zeta}=0.\tag{3.19}
\end{align}
After direct calculations, the above equation becomes
\begin{align}\notag
Y_{n,2}\left( \zeta \right) =\left( \begin{array}{c}
	\frac{c_+}{\varDelta _n}\\
	\zeta -r\\
\end{array} \right) +\sum_{j=1}^J{\frac{\left( \zeta -r \right) C_jk \left( \zeta _j \right) ^{-2n}Y_{n,1}\left( \zeta _j \right)}{\left( \zeta _j-r \right) \left( \zeta -\zeta _j \right)}}-\frac{1}{2\pi i}\int_{\left| \omega \right|=1}{\frac{\left( \zeta -r \right) k \left( \omega \right) ^{-2n+1}}{\left( \omega -\zeta \right) \left( \omega -r \right)}Y_{n,1}\left( \omega \right) \frac{\tau(\omega)}{\varsigma(\omega)}  d\omega}.\tag{3.20a}
\end{align}

Similarly, we consider the second element of Eq.~(3.14) now, divide it by $\zeta -\frac{1}{r}$ and remove the asymptotic behavior at $\zeta=0$ and the residue effect at zeros of $\bar{\varsigma}(\zeta)$, then we can get the expression of $Y_{n,1}(\zeta)$ as follows
\begin{align}
Y_{n,1}\left( \zeta \right) =\left( \begin{array}{c}
	r-\frac{1}{\zeta}\\
	-\frac{c_{+}}{\varDelta _n}\\
\end{array} \right) +\sum_{j=1}^{\bar{J}}{\frac{\left( \zeta -\frac{1}{r} \right) \bar{C}_jk \left( \bar{\zeta}_j \right) ^{2n}Y_{n,2}\left( \bar{\zeta}_j \right)}{\left( \bar{\zeta}_j-\frac{1}{r} \right) \left( \zeta -\bar{\zeta}_j \right)}}-\frac{1}{2\pi i}\int_{\left| \omega \right|=1}{\frac{\left( \zeta -\frac{1}{r} \right) k \left( \omega \right) ^{2n-1}}{\left( \omega -\zeta \right) \left( \omega -\frac{1}{r} \right)}Y_{n,2}\left( \omega \right) \frac{\bar{\tau}(\omega)}{\bar{\varsigma}(\omega)} d\omega}.\tag{3.20b}
\end{align}
Comparing Eq.~(3.20b) with the asymptotic behavior of $Y_{n,1}(\zeta)$ as $\zeta \rightarrow \infty$, one yields
\begin{align}
\frac{1}{\varDelta _n}=1+\sum_{j=1}^{\bar{J}}{\frac{\bar{C}_jk \left( \bar{\zeta}_j \right) ^{2n}Y_{n,2}^{\left( 1 \right)}\left( \bar{\zeta}_j \right)}{\left( r\bar{\zeta}_j-1 \right)}}-\frac{1}{2\pi i}\int_{\left| \omega \right|=1}{\frac{k \left( \omega \right) ^{2n-1}}{r\omega -1}Y^{\left( 1 \right)}_{n,2}\left( \omega \right) \frac{\bar{\tau}(\omega)}{\bar{\varsigma}(\omega)} d\omega},\tag{3.21a}
\end{align}
\begin{align}
\frac{-2bu_n-a}{\varDelta _n\sqrt{a^2+4b}}=-\frac{c_{+}}{\varDelta _n}+\sum_{j=1}^{\bar{J}}{\frac{\bar{C}_jk\left( \bar{\zeta}_j \right) ^{2n}Y^{\left( 2 \right)}_{n,2}\left( \bar{\zeta}_j \right)}{\left( \bar{\zeta}_j-\frac{1}{r} \right)}}-\frac{1}{2\pi i}\int_{\left| \omega \right|=1}{\frac{k \left( \omega \right) ^{2n-1}}{\omega -\frac{1}{r}}Y_{n,2}^{\left( 2 \right)}\left( \omega \right) \frac{\bar{\tau}(\omega)}{\bar{\varsigma}(\omega)} d\omega},\tag{3.21b}
\end{align}
which can be used to reconstruct the potential in the following section.

\vspace{5mm}\noindent\textbf{3.3 Time evolution and single soliton}

In what follows we consider the dependence of potentials and eigenfunctions on time $t$, rather than focusing solely on the influence of the spatial part, as in previous processes. As $n\rightarrow \infty$, Eq.~(1.2) can be reduced to $\frac{du_{\pm}}{dt}=0$, so potential is independent of $t$. As $n\rightarrow \infty$, we have
\begin{equation}\tag{3.22}
V_n\rightarrow \left( \begin{matrix}
	K_{n}^{\left( 1 \right)}\left( \lambda \right)&		L_{n}^{\left( 1 \right)}\left( \lambda \right)\\
	\sigma L_{n}^{\left( 1 \right)}\left( \lambda ^{-1} \right)&		K_{n}^{\left( 1 \right)}\left( \lambda ^{-1} \right)\\
\end{matrix} \right) +\left( \frac{a^2+4b}{4b} \right) ^2\left( \begin{matrix}
	K_{n}^{\left( 2 \right)}\left( \lambda \right)&		L_{n}^{\left( 2 \right)}\left( \lambda \right)\\
	\sigma L_{n}^{\left( 2 \right)}\left( \lambda ^{-1} \right)&		K_{n}^{\left( 2 \right)}\left( \lambda ^{-1} \right)\\
\end{matrix} \right)
\end{equation}
with
$$
K_{n}^{\left( 1 \right)}\rightarrow \frac{3a^2}{8b^2}\left( 4b\left( -1+\lambda ^2-c_{\pm}^{2} \right) +a^2\left( \lambda ^2-c_{\pm}^{2} \right) \right) ,
$$
$$
L_{n}^{\left( 1 \right)}\rightarrow \frac{3a^2}{8b^2}\left( a^2+4b \right) c_{\pm}\left( \lambda +\lambda ^{-1} \right) ,
$$
$$
K_{n}^{\left( 2 \right)}\rightarrow \frac{1}{4}\left( -6-3\lambda ^{-4}+\lambda ^4-4c_{\pm}^{2}\lambda ^2+12c_{\pm}^{4}+\left( 8+4c_{\pm}^{2} \right) \lambda ^{-2} \right) ,
$$
$$
L_{n}^{\left( 2 \right)}\rightarrow c_{\pm}\lambda ^{-3}\left( 1+\lambda ^2 \right) \left( 1+\lambda ^4-2\lambda ^2\left( 1+c_{\pm}^{2} \right) \right) .
$$
Combining above equation with Eqs.~(3.3), one obtain
\begin{equation}\tag{3.23}
\frac{\partial \psi _n}{\partial t}=p\psi _{n+1}+q\psi _n
\end{equation}
with
$$
p=\frac{1}{16b^2\lambda ^3}\left( a^2+4b \right) \left( 1+\lambda ^2 \right) \left( a^2\left( 1+\lambda ^4+\lambda ^2\left( 4-2c_{\pm}^{2} \right) \right) +4b\left( 1+\lambda ^4-2\lambda ^2\left( 1+c_{\pm}^{2} \right) \right) \right) ,
$$
\begin{align}\notag
q=&\frac{1}{16b^2\lambda ^4}\left( 16b^2\left( -\left( \lambda ^2-1 \right) ^2\left( 3+2\lambda ^2+3\lambda ^4 \right) +4c_{\pm}^{2}\left( \lambda +\lambda ^3 \right) ^2+12c_{\pm}^{4}\lambda ^4 \right) \right.\\ \notag
&\left.+a^4\left( -3+\lambda ^2\left( 4+4c_{\pm}^{2}+\lambda ^2\left( -26+4\lambda ^2-3\lambda ^4+4c_{\pm}^{2}\left( \lambda ^2-4 \right) +12c_{\pm}^{4} \right) \right) \right) \right.\\ \notag
&\left.+8a^2b\left( -3+\lambda ^2\left( 4+4c_{\pm}^{2}+\lambda ^2\left( -26+4\lambda ^2-3\lambda ^4+4c_{\pm}^{2}\left( \lambda ^2-1 \right) +12c_{\pm}^{4} \right) \right) \right) \right) .\notag
\end{align}
Here we assume the dependence of eigenfunctions on $t$ is manifested as
\begin{equation}\tag{3.24}
\hat{\varLambda} _n\left( t \right) =\left( \begin{matrix}
	e^{\kappa ^{\left( 1 \right)}t}&		0\\
	0&		e^{\kappa ^{\left( 2 \right)}t}\\
\end{matrix} \right) \varLambda _n\left( t \right) ,\ \ \hat{\varGamma} _n\left( t \right) =\left( \begin{matrix}
	e^{\kappa ^{\left( 1 \right)}t}&		0\\
	0&		e^{\kappa ^{\left( 2 \right)}t}\\
\end{matrix} \right) \varGamma _n\left( t \right) ,
\end{equation}
satisfying Eqs.~(3.4). Differentiating Eqs.~(3.24) with respect to $t$, then we have
$$
\frac{\partial \hat{\varGamma} _{n,1}\left( t \right)}{\partial t}=\kappa ^{\left( 1 \right)}\hat{\varGamma} _{n,1}\left( t \right) =p\hat{\varGamma} _{n+1,1}\left( t \right) +q\hat{\varGamma} _{n,1}\left( t \right) =\left( k^{-1}rp+q \right) \hat{\varGamma} _{n,1}\left( t \right) ,\ \ n\rightarrow +\infty ,
$$
$$
\frac{\partial \hat{\varLambda} _{n,2}\left( t \right)}{\partial t}=\kappa ^{\left( 2 \right)}\hat{\varLambda} _{n,2}\left( t \right) =p\hat{\varLambda} _{n+1,2}\left( t \right) +q\hat{\varLambda} _{n,2}\left( t \right) =\left( krp+q \right) \hat{\varLambda} _{n,2}\left( t \right) ,\ \ n\rightarrow -\infty ,
$$
which can be deduced to
\begin{equation}\tag{3.25}
\kappa ^{\left( 1 \right)}=k^{-1}rp+q,\ \ \kappa ^{\left( 2 \right)}=k\,rp+q.
\end{equation}
Introducing $
\varTheta =\left( \begin{matrix}
	\kappa ^{\left( 1 \right)}&		0\\
	0&		\kappa ^{\left( 2 \right)}\\
\end{matrix} \right) $, the dependence of the Jost eigenfunctions on time can be expressed as
$$
\frac{\partial \varLambda _n}{\partial t}=V_n\varLambda _n-\varLambda _n\varTheta ,\ \ \frac{\partial \varGamma _n}{\partial t}=V_n\varGamma _n-\varGamma _n\varTheta ,
$$
right now. Recalling that matrix $T(\zeta)$ can express the linear correlation between two Jost eigenfunctions, the time evolution of scattering coefficient can be easily obtained as follows
\begin{equation}\tag{3.26}
\frac{\partial \varTheta}{\partial t}=\varTheta T-T\varTheta =\left( \begin{matrix}
	0&		\bar{\tau}\left( \kappa ^{\left( 2 \right)}-\kappa ^{\left( 1 \right)} \right)\\
	\tau \left( \kappa ^{\left( 1 \right)}-\kappa ^{\left( 2 \right)} \right)&		0\\
\end{matrix} \right) ,
\end{equation}
\begin{equation}\tag{3.27}
\varsigma \left( t \right) =\varsigma \left( 0 \right) ,\ \ \bar{\varsigma}\left( t \right) =\bar{\varsigma}\left( 0 \right) .
\end{equation}
For the discrete eigenvalue $\lambda_j$, one have $
\varLambda _{n,2}\left( \lambda _j,t \right) =b_j\varGamma _{n,1}\left( \lambda _j,t \right)$, then the norming constant can be rewritten as
\begin{equation}\tag{3.28a}
\tau _j\left( t \right) =\tau _j\left( 0 \right) e^{\left( \kappa ^{\left( 1 \right)}\left( \lambda _j \right) -\kappa ^{\left( 2 \right)}\left( \lambda _j \right) \right) t},\ \ \bar{\tau}_j\left( t \right) =\bar{\tau}_j\left( 0 \right) e^{\left( \kappa ^{\left( 2 \right)}\left( \bar{\lambda}_j \right) -\kappa ^{\left( 1 \right)}\left( \bar{\lambda}_j \right) \right) t},
\end{equation}
\begin{equation}\tag{3.28b}
C_j\left( t \right) =C_j\left( 0 \right) e^{\left( \kappa ^{\left( 1 \right)}\left( \lambda _j \right) -\kappa ^{\left( 2 \right)}\left( \lambda _j \right) \right) t},\ \ \bar{C}_j\left( t \right) =\bar{C}_j\left( 0 \right) e^{\left( \kappa ^{\left( 2 \right)}\left( \bar{\lambda}_j \right) -\kappa ^{\left( 1 \right)}\left( \bar{\lambda}_j \right) \right) t}.
\end{equation}

Finally, we study the reflectionless case with one pair of eigenvalues $\zeta _1$, $\bar{\zeta}_1=\zeta _{1}^{-1}$ (see Fig.~8). In this case, noting that $b$ must be greater than $-\frac{a^2}{4}$ and the potential
\begin{equation}\tag{3.29}
u_n=\frac{\sqrt{a^2-4b}}{2b}\left( c_+-\frac{\bar{C}_1k\left( \bar{\zeta}_1 \right) ^{2n}}{\left( \bar{\zeta}_1-\frac{1}{r} \right)}Y_{n,2}^{\left( 2 \right)}\left( \bar{\zeta}_1 \right) \varDelta _n \right) -\frac{a}{2b},
\end{equation}
in which unknown variables are determined by the following system of equations
\begin{equation}\tag{3.30}
\left\{ \begin{array}{l}
	Y_{n,21}\left( \bar{\zeta}_1 \right) =\frac{c_+}{\varDelta _n}+\frac{\left( \bar{\zeta}_1-r \right) C_1k^{-2n}}{\left( \zeta _1-r \right) \left( \bar{\zeta}_1-\zeta _1 \right)}Y_{n,11}\left( \zeta _1 \right) ,\\
	Y_{n,22}\left( \bar{\zeta}_1 \right) =\bar{\zeta}_1-r+\frac{\left( \bar{\zeta}_1-r \right) C_1k^{-2n}}{\left( \zeta _1-r \right) \left( \bar{\zeta}_1-\zeta _1 \right)}Y_{n,12}\left( \zeta _1 \right) ,\\
	Y_{n,11}\left( \zeta _1 \right) =r-\zeta _{1}^{-1}+\frac{\left( \zeta _1-r^{-1} \right) \bar{C}_1k^{2n}}{\left( \bar{\zeta}_1-r^{-1} \right) \left( \zeta _1-\bar{\zeta}_1 \right)}Y_{n,21}\left( \bar{\zeta}_1 \right) ,\\
	Y_{n,12}\left( \zeta _1 \right) =-\frac{c_+}{\varDelta _n}+\frac{\left( \zeta _1-r^{-1} \right) \bar{C}_1k^{2n}}{\left( \bar{\zeta}_1-r^{-1} \right) \left( \zeta _1-\bar{\zeta}_1 \right)}Y_{n,22}\left( \bar{\zeta}_1 \right) ,\\
	\frac{1}{\varDelta _n}=1-\frac{C_1k^{-2n}}{\left( \zeta _1-r \right) \zeta _1}Y_{n,12}\left( \zeta _1 \right) .\\
\end{array} \right.
\end{equation}
\begin{center}
\includegraphics[scale=0.5]{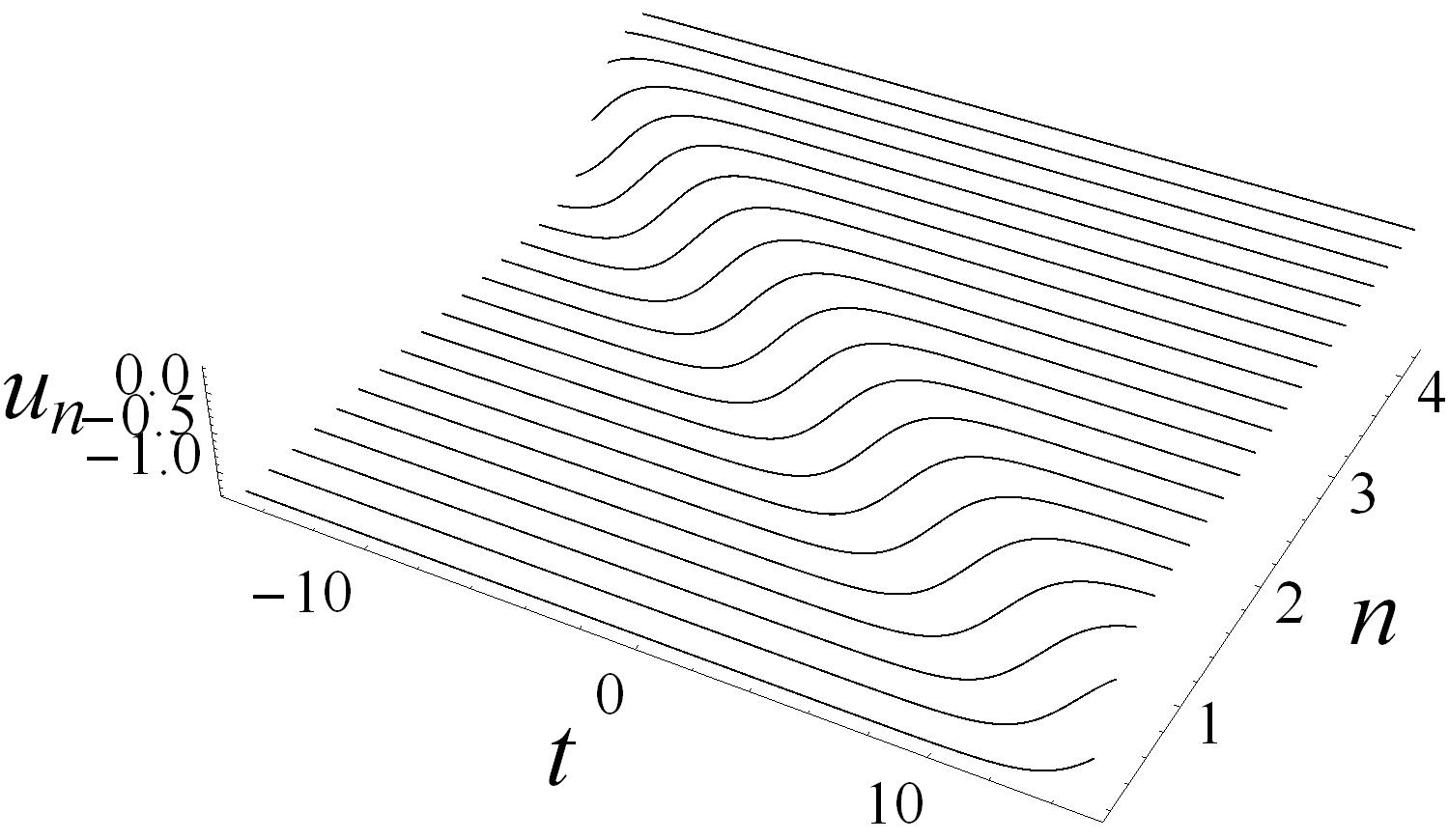}\hspace{2.3cm}
\includegraphics[scale=0.5]{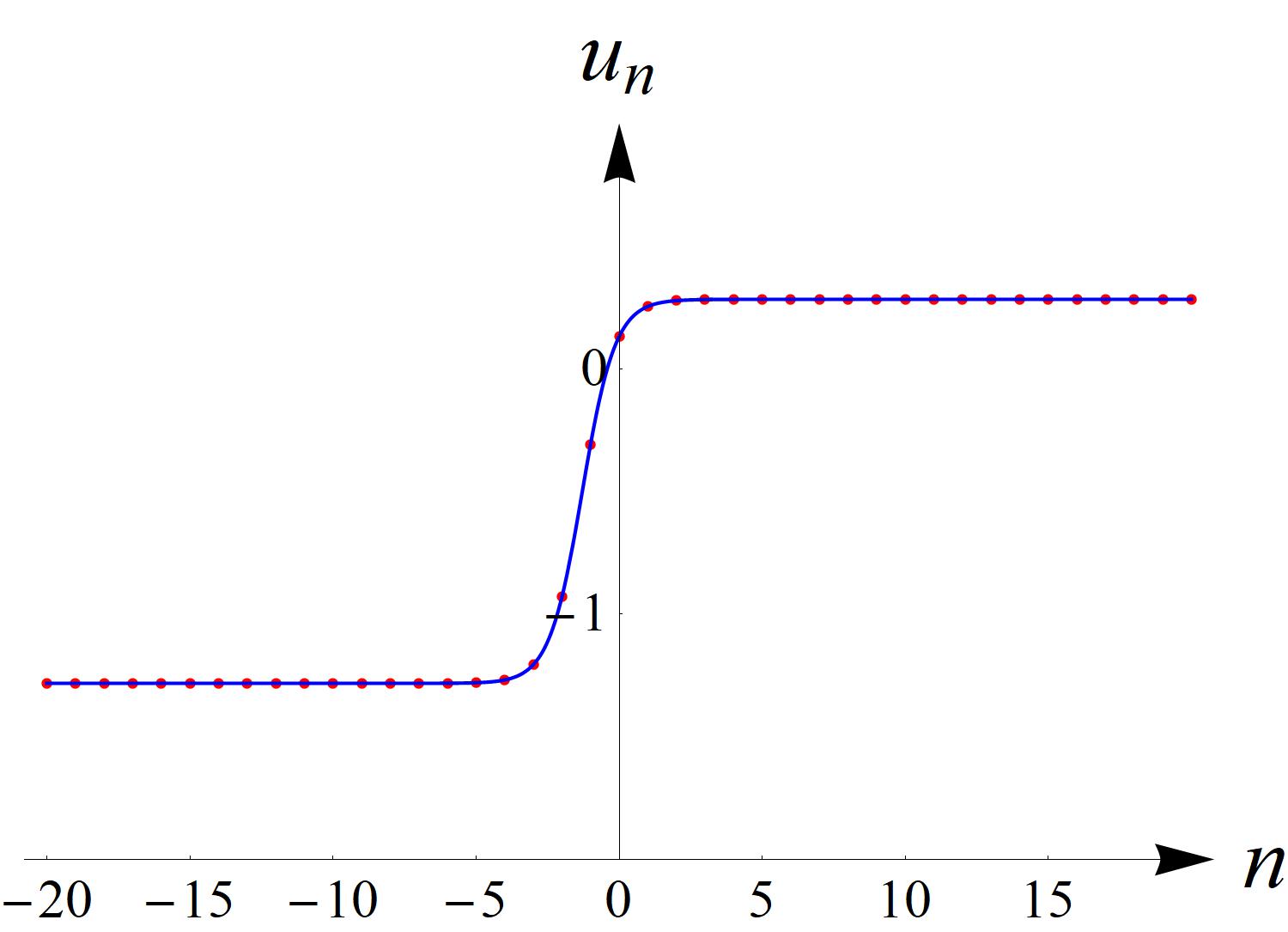}

\vspace{-0.2cm}{\footnotesize\hspace{0.5cm}(a)\hspace{7.5cm}(b)}\\
\flushleft{\footnotesize
\textbf{Fig.~$8$.} The single soliton solution corresponding to Eq.~(3.29) under a special step-like boundary condition with $a=1$, $b=1$, $c_0=0.7$, $\bar{\zeta}_1=\frac{1-c_0}{r}$, $C_1(0)=0.5$.}
\end{center}

From Eqs.~(3.29)(3.30), we can derive the velocity for the kink in terms of
\begin{equation}\tag{3.31}
V_{kink}=\frac{\kappa ^{\left( 2 \right)}\left( \bar{\lambda}_j \right) -\kappa ^{\left( 1 \right)}\left( \bar{\lambda}_j \right)}{\log \left( k^2\left( \lambda _j \right) \right)}.
\end{equation}
For the discrete eigenvalue in Fig.~8, the velocity is reduced to $\frac{\left( a^2+4b \right) c_0\left( 3a^2-\left( a^2+4b \right) c_{0}^{2} \right)}{2b^2\log \left( \frac{1-c_0}{1+c_0} \right)}$ depending on the values of $a$, $b$ and $c_0$.

\vspace{5mm}\noindent\textbf{4  Conclusions}\\
\hspace*{\parindent}
In this work, we have studied two kinds of boundary conditions for the semi-discrete Gardner equation. It has been proved that the semi-discrete Gardner equation has similar properties to its continuous counterpart including the existence of the kink solutions, soliton solutions with different polarities and so-called rogue wave solutions, in addition, the polarity of solitons has an important influence on the collision results of two solitons.

For the first boundary condition, the meromorphic functions in inverse problem have high-order poles. Firstly, we have discussed the case of one-order poles, for which there exist the single solitons with different polarities. When there are two sets of discrete eigenvalues, we have considered the soliton-soliton interaction problem, and analyzed how the parameters $a$, $b$ affect the types of collision, including the head-on collision and the overtaking collision. Further, we have found that if the polarities of the two solitons are opposite, the energy at the moment of collision increases, and the amplitude is more than twice that of the original soliton, generating the so-called rogue wave phenomenon. Considering the double zeros of the scattering coefficients, we have constructed the two-order poles solution of the semi-discrete Gardner equation, which shows the interaction of a positive amplitude wave and a negative one, however, it does not produce large amplitude at the moment of collision as the two-soliton collision in Section 2.3.

In addition, a specific step-like boundary condition can evolve kink solutions for the semi-discrete Gardner equation, which have been confirmed by solving the reflectionless potential using the inverse scattering transform in Section 3. We have derived the propagation velocity of the kink and found that it depends on the parameters $a$, $b$ in the semi-discrete Gardner equation and the absolute value of the boundary amplitude for the fixed discrete eigenvalue. As an undercompressive dispersive shock wave, Whitham modulation theory is generally an effective method to study the kink. It is worth discussing this content in the semi-discrete Gardner equation, which is beyond the scope of this work.

In summary, all the work in this paper is carried out in the framework of the inverse scattering transform with a zero reflection. On the contrary, the non-zero reflection coefficient makes the continuous spectrum contribute to the reconstruction potential, which is usually manifested as the radiation. For the semi-discrete Gardner equation, how the solution with a non-zero reflection evolves needs to be strictly analyzed by long-time asymptotic analysis.

\vspace{5mm}\noindent\textbf{Acknowledgments}\\
\hspace*{\parindent}

We express our sincere thanks to each member of our discussion group for their suggestions. This work has been supported by the National Natural Science Foundation of China under Grant No. 12472387, National Natural Science Foundation of China under Grant No. 11905155, and the Fund Program for the Scientific Activities of Selected Returned Overseas Scholars in Shanxi Province under Grant No. 20220008.

\end{document}